%% file: architecture_judgment.tex
\documentclass[manuscript,screen]{acmart}
\AtBeginDocument{%
  }

\usepackage{subcaption}
\usepackage{cleveref}
\usepackage{xspace}
\usepackage{tikz}
\usepackage{tabularx}
\usepackage{multicol}
\usepackage[most]{tcolorbox}

\definecolor{quoteaccent}{HTML}{2C6E9B} % accent bar + speaker tags
\definecolor{quotebg}{HTML}{F4F7FA}     % soft tinted background

\newcommand{\pid}[1]{%
  \tcbox[on line, boxsep=1pt, left=4pt, right=4pt, top=1pt, bottom=1pt,
         arc=3pt, colback=quoteaccent, colframe=quoteaccent,
         fontupper=\sffamily\bfseries\scriptsize\color{white}]{#1}}

\newcommand{\iquote}[2]{\textit{``#2''}\nobreakspace\pid{#1}}

\newtcolorbox{pquote}[1]{%
  enhanced, unbreakable, sharp corners,
  colback=quotebg, colframe=quotebg, boxrule=0pt,
  borderline west={3pt}{0pt}{quoteaccent},
  left=10pt, right=10pt, top=5pt, bottom=5pt,
  before skip=4pt, after skip=8pt,
  fontupper=\itshape,after={\par\noindent}, 
  after upper={\nobreakspace\normalfont\pid{#1}}%
}

\newtcolorbox{pquotes}{%
  enhanced, unbreakable, sharp corners,
  colback=quotebg, colframe=quotebg, boxrule=0pt,
  borderline west={3pt}{0pt}{quoteaccent},
  left=10pt, right=10pt, top=5pt, bottom=5pt,
  before skip=4pt, after skip=8pt,after={\par\noindent}
}
\newcommand{\utt}[2]{%
  \par\noindent\hangindent=2.4em\hangafter=1
  \pid{#1}\hspace{0.6em}\textit{#2}\par\vspace{4pt}%
}

\definecolor{notebg}{HTML}{FFF6CC}    % pale sticky-note yellow
\definecolor{notecited}{HTML}{FFE38A} % darker yellow for cited metaphors
\newtcolorbox{metaphorbox}[1]{%
  enhanced, sharp corners,
  colback=#1, colframe=#1, boxrule=0pt,
  drop fuzzy shadow=black!35,
  before skip=0pt, after skip=8pt,
  left=5pt, right=5pt, top=4pt, bottom=3pt, middle=3pt,
  segmentation hidden,
  fontupper=\footnotesize\itshape, before upper=\raggedright%
}
\NewDocumentCommand{\metaphor}{s m m}{%
  \begin{metaphorbox}{\IfBooleanTF{#1}{notecited}{notebg}}#3\tcblower\pid{#2}\end{metaphorbox}}

\definecolor{cocaccent}{HTML}{1F7A78} % teal title bar (complements quote blue)
\definecolor{cocbg}{HTML}{EEF6F5}     % soft teal-tinted background
\definecolor{cocrule}{HTML}{124C4A}   % deep teal west stripe

\newtcolorbox{cocbox}[1][Central Organising Concept]{%
  enhanced, unbreakable, sharp corners,
  colback=cocbg, colframe=cocaccent, boxrule=0pt,
  borderline west={3pt}{0pt}{cocrule},
  left=10pt, right=10pt, top=8pt, bottom=8pt,
  before skip=8pt, after skip=8pt,
  colbacktitle=cocaccent, coltitle=white,
  fonttitle=\sffamily\bfseries\small,
  lefttitle=10pt, toptitle=3pt, bottomtitle=3pt,
  title={#1},
  fontupper=\itshape,
}
\newcommand{\coc}[2][]{%
  \def\coctheme{#1}%
  \begin{cocbox}[Central Organising Concept%
    \ifx\coctheme\empty\else\space\textmd{\textbf{·}}\space\textmd{#1}\fi]%
  #2%
  \end{cocbox}%
}

\newtcolorbox{questionbox}{%
  enhanced, breakable, sharp corners,
  colback=cocbg, colframe=cocbg, boxrule=0pt,
  borderline west={3pt}{0pt}{cocrule},
  left=10pt, right=10pt, top=5pt, bottom=5pt,
  before skip=6pt, after skip=6pt,
  fontupper=\itshape,
}
\newcommand{\framingquestion}[1]{%
  \begin{questionbox}#1\end{questionbox}%
}

\definecolor{revaccent}{HTML}{9A5B13} % ochre title bar + labels
\definecolor{revbg}{HTML}{FDF4E8}     % warm tinted background
\definecolor{revrule}{HTML}{6B3E0B}   % deep ochre west stripe

\newif\ifrevresponses \revresponsestrue

\newcounter{revresponse}
\crefname{revresponse}{response}{responses}
\Crefname{revresponse}{Response}{Responses}

\newcommand{\revlabel}[1]{%
  \tcbox[on line, boxsep=1pt, left=4pt, right=4pt, top=1pt, bottom=1pt,
         arc=3pt, colback=revaccent, colframe=revaccent,
         fontupper=\sffamily\bfseries\scriptsize\color{white}]{#1}}

\newtcolorbox{revresponsebox}[1][Reviewer Request \& Response]{%
  enhanced, breakable, sharp corners,
  colback=revbg, colframe=revaccent, boxrule=0pt,
  borderline west={3pt}{0pt}{revrule},
  left=10pt, right=10pt, top=8pt, bottom=8pt,
  before skip=8pt, after skip=8pt,
  colbacktitle=revaccent, coltitle=white,
  fonttitle=\sffamily\bfseries\small,
  lefttitle=10pt, toptitle=3pt, bottomtitle=3pt,
  title={#1},
  fontupper=\sffamily\small,   % sans-serif body: clearly not running text
  after={\par\noindent}%
}

\newlength{\revhang}

\newcommand{\revresponse}[3][]{%
  \ifrevresponses
    \refstepcounter{revresponse}%
    \def\revsource{#1}%
    \begin{revresponsebox}[Comment \& Response~\therevresponse
      \ifx\revsource\empty\else\space\textmd{\textbf{·}}\space\textmd{#1}\fi]%
      \settowidth{\revhang}{\revlabel{Comment}\hspace{0.6em}}%
      \par\noindent\hangindent=\revhang\hangafter=1
      \makebox[\revhang][l]{\revlabel{Comment}}\textit{#2}\par
      \vspace{5pt}%
      \par\noindent\hangindent=\revhang\hangafter=1
      \makebox[\revhang][l]{\revlabel{Response}}#3\par
    \end{revresponsebox}%
  \fi
}

\begin{document}

%% ---------------------------------------------------------------------------
%% Draft revision marking
%% ---------------------------------------------------------------------------
%% \draft{<rev>}{<text>} marks <text> as belonging to revision <rev>.
%%   * <rev> greater than the `revision' counter below  -> typeset in blue,
%%     i.e. the text was added or changed after that revision.
%%   * <rev> equal to or lower than the counter          -> typeset normally.
%% Bump \setcounter{revision}{...} once a round of changes has been read and
%% accepted; everything at or below that number then folds back into the prose.
%% Safe mid-sentence: the macro adds no space of its own on either side, so
%% `a \draft{2}{b} c' and `a\draft{2}{b}c' both behave exactly like the text
%% written out by hand.
\newcounter{revision}
\setcounter{revision}{8}

\definecolor{draftblue}{HTML}{0066FF} % bright, fully saturated blue

\DeclareRobustCommand{\draft}[2]{%
  \ifnum#1>\value{revision}\relax%
    \begingroup\color{draftblue}#2\endgroup%
  \else%
    #2%
  \fi%
}

%% The old author-note macro (red italics), unchanged apart from its name.
\newcommand{\draftnote}[1]{\emph{\textcolor{red}{#1}}}

\newcommand{\university}{\emph{University of Victoria}}
\newcommand{\numstudents}{X\xspace}


\title{Surrendering and Reclaiming Judgment: How Software Engineering Students Experience AI Reliance}

 \author{Nathan Cassee}
 \email{nathancassee@uvic.ca}
 \affiliation{
    \institution{University of Victoria}
    \country{Canada}
 }

  \author{Umit Akirmak}
 \email{uakirmak@uvic.ca}
 \affiliation{
    \institution{University of Victoria}
    \country{Canada}
 }

  \author{Luiz Guerra}
 \email{luizguerra@uvic.ca}
 \affiliation{
    \institution{University of Victoria}
    \country{Canada}
 }

   \author{Margaret-Anne Storey}
 \email{mstorey@uvic.ca}
 \affiliation{
    \institution{University of Victoria}
    \country{Canada}
 }

  \author{Neil Ernst}
 \email{nernst@uvic.ca}
 \affiliation{
    \institution{University of Victoria}
    \country{Canada}
 }

\begin{abstract}
  \draft{5}{Software architecture is one of the skills novice developers must master to succeed in industry, and as AI disrupts software development, it is also transforming how students learn architecture. Existing work on how students and novices work with AI has mostly focused on small programming exercises or on students' perceptions of AI, so little is known about how students experience AI \draft{7}{as they learn software architecture in larger team projects}.
  We report on an exploratory case study of 43 senior software engineering students who used AI throughout a semester-long, team-based software architecture project. We combined surveys with focus groups and analyzed the focus group transcripts and the open survey responses through Braun \& Clarke's reflexive thematic analysis.
  We constructed two themes. The first, \textbf{Surrendering Judgment to AI}, traces how participants' reliance on AI deepened as they used it to architect and implement their systems, until they surrendered their judgment and became stuck in a codebase they had created but could not understand, which made them rely on AI even more. Participants described feeling ashamed, frustrated, and emotionally disconnected from their work, a toll that, in our reading, left them with even less reason to break their reliance. The second, \textbf{Applying Human Judgment}, centers on participants who kept or reclaimed their judgment: the knowledge and experience it rested on, and the strategies they used to apply it. Participants' accounts suggest that relaxed deadlines and a lecture and lab on working with AI gave them room to acquire knowledge and reclaim judgment.
  Given the expectations placed on junior developers and the widespread adoption of AI, we believe these first-hand accounts can help software engineering education and industry adopt AI in ways that preserve novice developers' judgment, ownership, and motivation.}
\end{abstract}

 \keywords{Software Engineering Education, Experience of Generative AI Reliance, Emotional Disconnect}


\maketitle

\input{sections/1-introduction}

\input{sections/3-related-work}

\input{sections/4-methodology}
\input{sections/5-results}

\input{sections/6-discussion}
\input{sections/7-limitations}
\input{sections/8-conclusion}

\bibliographystyle{ACM-Reference-Format}
\bibliography{references}

\end{document}

%% file: sections/1-introduction.tex
\section{Introduction}

Competent and productive developers are expected to master a variety of higher-order skills. One of the most important skills is software architecture, or the ability to design software architectures~\cite{Hasselbring:2018}. Mastering architecture is particularly difficult for novices, because its open-ended and complex nature makes it hard to learn~\cite{Galster:2016}. Therefore, novices need sustained practice: repeated engagement with realistic, difficult software architecture problems~\cite{Deursen:2017, Loksa:2016}.

The widespread adoption of generative AI in software engineering\footnote{\url{https://survey.stackoverflow.co/2025}} might disrupt how novices master software architecture, as AI can reason about software architecture~\cite{Guerra:2025} and even suggest entire architectures~\cite{Esposito:2026}. These capabilities make AI very useful for experienced developers and may increase their productivity~\cite{Schmid:2025, Ullrich:2025}. However, how these benefits extend to novices is less clear, because a developer's experience affects how they use AI for development~\cite{Jaman:2026}, and the capabilities that enable experienced developers to use AI for architecture might not serve novices in the same way. In fact, it might be attractive for novices who have not yet mastered architecture to offload the demanding parts of architectural work to AI~\cite{Choudhuri:2026}. 

At the same time, software engineering education is rapidly exploring how to incorporate generative AI~\cite{Fenu:2024, Rahe:2025}. Educators are redesigning their courses, architecture courses in particular, with the expectation that AI can take over routine implementation work. This should free students to spend their time on the higher-order architecture skills these courses are meant to teach~\cite{Groothuijsen:2024, Salomon:2025}. Because AI is new and rapidly evolving, there are no established guidelines on how it should be adopted. \draft{1}{Existing work that might inform guidelines has focused mostly on foundational skills like programming~\cite{Fenu:2024, Aashish:2024, Oliveira:2025, Rahe:2025}, in individual, short-lived exercises, and it characterizes primarily \emph{how} novices use AI~\cite{Guner:2025,Margulieux:2024,Rojas-Galeano:2025,Salomon:2025} through self-reported perceptions that may not match what happens during development~\cite{Butler:2025, METR:2025:Productivity}.} However, this work does not focus on what it's like for novice developers to work with AI while learning challenging skills like software architecture in a semester-long, team-based project. \draft{6}{We therefore ask: \textbf{How do senior software engineering students experience AI as they learn software architecture in a team project?}}

We conducted an exploratory case study of 43 senior software engineering students who used AI throughout a semester-long, team-based software architecture project. We \draft{5}{combined} three surveys with four focus groups, and we analyzed \draft{5}{the focus group transcripts and the open survey responses} through reflexive thematic analysis~\cite{Braun:2022}.

\draft{5}{In this paper, we report on the two themes we constructed through reflexive thematic analysis. Both are about what it was like for students to work with AI, and both center on participants' judgment over what AI produced. Our first theme, \textbf{Surrendering Judgment to AI}, is constructed around participants' accounts of an extensive reliance on AI, of surrendering their judgment, first over architectural decisions and then over code they could no longer evaluate, and of the emotional toll that followed: they felt ashamed that they had to rely on AI, frustrated when incorrect information from AI left them stuck, and increasingly disconnected from their work as their reliance deepened. One participant described AI as an ``intelligent gaslighter'', and another as ``that friend that you meet and think he's really smart [...] And he's been manipulating you''.}

\draft{5}{Our second theme, \textbf{Applying Human Judgment}, centers on \draft{8}{the accounts of participants who spoke about applying} their judgment while working with AI, and on participants who reclaimed it during the course. What stood out to us is that their judgment was anchored in software engineering knowledge: some brought it into the project, and others recognized during the project that they lacked it and acquired it. Participants who reclaimed their judgment also described changing how they worked with AI. One of them put it like this: ``AI was like milk and a cup of tea. A little bit would enhance the drink, but too much could ruin it and require you to remake it.''}

\draft{5}{To us, these results suggest that the current generation of students and novice developers may be particularly vulnerable. In participants' accounts, three conditions pushed them toward surrender: an unfamiliar technology stack, delivery pressure outside their control, and an expectation that they use AI. None of these is unique to a course project, as junior developers entering industry experience all three~\cite{Xin:2024, Begel:2008, Miller:2025}. We believe, these conditions can contribute to a sustained reliance on AI that erodes the motivation to learn, and without that motivation there is little reason to break the reliance. Our second theme, applying human judgment, suggests this outcome is not inevitable: participants told us how relaxed deadlines and a lecture and lab on working with AI helped some of them reclaim their judgment. These changes hint at interventions that can help developers and organizations adopt AI sustainably, in ways that preserve novice developers' judgment, ownership, and motivation.}

\draft{5}{This paper makes three contributions. First, \draft{7}{we show that what participants surrendered to AI was their judgment over what AI was doing. \citet{Storey:2026} describes cognitive surrender by what developers stop doing (\Cref{sec:related-work}); our participants' accounts help us understand what they gave up. In participants' own words, we also document what that surrender felt like, including an emotional toll that, to our knowledge, prior work has not reported. Read through Self-Determination Theory~\cite{Ryan:2000}, these accounts suggest why surrender left participants disconnected from their work (\Cref{sec:discussion:sdt}). Deadlines they could not control undermined their sense of autonomy, and working beside an AI they felt knew more than they did undermined their sense of competence; together, we think this contributed to their work feeling hollow}.}

\draft{7}{Second, we show what the judgment of participants who kept or reclaimed it rested on: knowledge of the technologies AI worked with and strategies for working with AI that they discovered during the project. Participants started a new chat when AI went off track, kept each chat scoped to the problem at hand, wrote custom instructions, had AI work in small steps they could test, edited the plans AI proposed before it wrote any code, and asked AI to explain the code it produced. Each strategy helped participants apply judgment, gave them the chance to notice when AI made mistakes, and helped preserve their own understanding of the code.}

\draft{7}{Third, we argue that instruments that can detect surrender are necessary. Existing closed survey items did not allow us to detect the surrender; only participants' own words, in the open survey responses and the focus groups, allowed us to describe what surrender was like for these participants. We believe these accounts can be used by educators and researchers to help identify surrender in their own courses, or to develop instruments that can help detect it.}

%% file: sections/3-related-work.tex
\section{Related Work}
\label{sec:related-work}

Our study focuses on the experiences of senior software engineering students as they adopt generative AI on a software architecture project. We position our contributions in light of research on generative AI in software engineering education, organized by what that literature asks: what students \emph{do} with AI, what they \emph{think} of it, what relying on it \emph{costs} them, and, in a much smaller body of work, what using it is \emph{like} for them. Almost none of this work concerns software architecture education.

\paragraph{Introductory programming}
Most of what is known about students and generative AI comes from foundational programming courses~\cite{Fenu:2024,Aashish:2024,Rahe:2025,Guner:2025,Margulieux:2024,Rojas-Galeano:2025,Groothuijsen:2024}. These studies catalog how students use AI on their coursework and consistently report wide variation: some students have AI produce entire assignments, while others use it only to look things up or make small changes. Their participants, however, are first- and second-year students working individually on short, well-specified exercises~\cite{Fenu:2024,Rahe:2025,Guner:2025}, a poor proxy for the open-ended, team-based work of software architecture that we study.

\paragraph{Team projects}
Fewer studies observe students while they're working on longer team projects, and the studies that do catalog what changed in how students work, staying close to participants' explicit, surface-level accounts rather than interpreting what the experience was like for them. \citet{Salomon:2025} surveyed 349 students and interviewed 24 of them in a project-based course where teams built three versions of a system with unrestricted access to AI. They find that students used AI not only to generate code but to validate work retrospectively against requirements and design, consulted AI to find information about software design, and that they turned to AI before consulting teammates, which reduced direct collaboration. 
\citet{Mircea:2026} surveyed 150 students across 18 capstone teams, finding that voluntary AI use was near-universal across the lifecycle, and that students articulated their own limits, recommending verification and warning against losing independent understanding. \citet{Wang:2026} examine a setting much like ours: a term-long, team-based greenfield project in which students directed their own use of generative AI. Asking which skills should be taught as generative AI becomes common in software development, they find that non-programming skills, such as working out requirements and specifications, matter more for effective human--AI collaboration than they used to.

\paragraph{Software architecture}
Software architecture itself remains barely studied. \citet{Ferino:2026} reviewed 80 studies of novices adopting LLMs, and found only three that report novices using LLMs for architectural work: one on architecture definition, and two on diagram generation. They call this small number a research gap, and name architecture as a direction for future research. \citet{Mitas:2025} address that gap from the tool side, proposing an AI-driven tutor for cloud-based software architectures.

\paragraph{Perceptions of AI}
\draft{7}{Beyond how students use AI, related work studies what they think of it, and most of these studies report a mix of positive and negative views~\cite{Ferino:2026}. Views split most clearly on learning. Most students in an undergraduate Java course judged ChatGPT suitable for learning programming concepts, but some avoided it for fear of not developing their own proficiency~\cite{Haindl:2024}. First-year engineering students listed over-reliance that impairs learning among their concerns~\cite{Mohammed:2025}, and computing students at one university grew more worried about becoming over-reliant on AI across three survey rounds~\cite{Keuning:2024}. Views also split by task: students judge AI more useful for coding and debugging than for analysis and design~\cite{Baresi:2025}, the work an architecture course centers on.} This literature records feeling as well as opinion: \citet{Ferino:2026} report that 56 of the 80 studies they reviewed capture novices' emotions towards LLM tools, including satisfaction, surprise, frustration, and fear. What it captures, however, is sentiment aimed at the tools and belief about them, not what students encountered in doing the work, which is the focus of our study.

\paragraph{Cognitive engagement}
Closest to our study is work that focuses on the consequences of AI use and how those consequences are experienced by students. Applying structural equation modeling to survey data from 299 STEM students across five North American universities, \citet{Choudhuri:2026} find that students who trusted and routinely used generative AI reported having significantly lower cognitive engagement, and that neither prior experience with AI nor time in academia protected them. However, this work \draft{7}{measures engagement through closed survey items, so it cannot capture, in students' own words, what working with AI was like}. \draft{7}{\citet{Storey:2026} has begun to describe a related pattern among developers, ``cognitive surrender'' to AI: adopting its output with minimal scrutiny, as opposed to cognitive offloading, the strategic delegation of a discrete task to a tool. Cognitive surrender describes what developers stop doing when they surrender. It does not say what they give up, or what surrender is like for the developer who experiences it.}

\paragraph{Students' experience of AI use}
\draft{5}{Three studies} report the encounter of students with AI more directly. \citet{Choudhuri:2024} ran a between-subjects study in which 22 students completed three short software engineering tasks. Students using ChatGPT were no more productive, and felt no more capable, than students using traditional resources, but they were significantly more frustrated; tracing five AI faults to seven consequences, the authors report uncertainty about correctness and self-doubt among them. \citet{Anderson:2026:Stuck} had ten advanced students make a non-trivial change to an unfamiliar codebase over three hours, half of them with access to ChatGPT, and nine of the ten became stuck at least once. They contribute a typology of seven kinds of stuck moment and of how AI helped, hindered, or caused each impasse; several of these mechanisms recur in our data. They also find that task success and system comprehension come apart, as AI allowed participants with only a medium level of comprehension to complete a task that comparable participants without AI did not, and they close by calling for study of the long-term implications of working with a reduced understanding of the system. \draft{5}{\citet{Tarek:2026} used a design similar to that of \citet{Anderson:2026:Stuck}, and proposes a taxonomy of personas that describes how novice programmers rely on AI: self-sufficient developers, understanding-gated developers, and AI-steered developers. These personas distinguish modes of working with AI, but not what it was like for students to program with AI, or to rely on it.}

\citet{Choudhuri:2025:Trenches} used reflexive thematic analysis on interviews with sixteen students drawn from senior and capstone software engineering courses. They find that students use AI across a wide range of learning and implementation tasks, and that some describe careful practices, such as cross-checking AI answers against documentation and Stack Overflow, though the authors stress that over-reliance and initial blind trust remain common among beginners. Several of the challenges they report echo what we find: students struggling to make sense of misleading responses, and struggling to get AI to do what they intend. Their account, however, organizes those challenges by where AI helps and hurts across task phases and traces them to properties of the tools, rather than grounding them in how a reliance on AI made participants feel about themselves and their work.

\paragraph{Research gap}
\draft{7}{Taken together, prior work tells us what students do with AI, what they believe about it, and what relying on it might cost them. It does not focus as much on what working with AI is like for the students themselves while they learn a skill as demanding as software architecture. This question is important because courses are rapidly adopting AI, based on the expectation that students will use AI~\cite{Groothuijsen:2024,Salomon:2025}, yet there are no established guidelines on how they should. Students in these courses must learn a demanding skill while working with a tool that can do much of that work for them. We therefore argue that it is essential to capture not only how students use AI, but also how they experience working with it under these conditions: what relying on it does to their judgment over their work and to their motivation for it, and how they work with it to preserve their judgment.}

\draft{7}{Existing work rarely asks this. Studies with experienced developers measure productivity~\cite{Schmid:2025,Ullrich:2025}, and studies in education characterize how novices use AI, mostly through self-reported perceptions~\cite{Guner:2025,Margulieux:2024,Rojas-Galeano:2025,Salomon:2025}.} Studies that focus on the experience of working with AI compress it into \draft{5}{at most} a few hours of individual work on a prescribed task\draft{5}{~\cite{Choudhuri:2024,Anderson:2026:Stuck,Tarek:2026}}. Those that span a whole project ask about workflow, collaboration, or skills rather than experience~\cite{Salomon:2025,Mircea:2026,Wang:2026}, and none study software architecture~\cite{Ferino:2026}. The work that does focus on experiential accounts~\cite{Choudhuri:2025:Trenches} does not situate those accounts in the context of a single project, nor does it explore the emotional impact of AI reliance. \citet{Ferino:2026} arrive at a similar gap from the other direction: only 19 of their 80 studies consider motivation, and those studies report why novices take up an LLM tool or refuse it, not what a sustained reliance does to their motivation for the work. They also name the effect of LLM tools on novice developers' mental health as an open research gap.

We address this \draft{5}{gap} by following a single cohort through one semester-long, team-based software architecture project, combining surveys administered as the project unfolded with focus groups held after its final phase. \draft{5}{We construct an interpretive account of how these senior software engineering students experienced a deepening reliance on AI, including the emptiness and loss of motivation some described, and of how other students kept or reclaimed their judgment.}

%% file: sections/4-methodology.tex
\section{Methodology}
\label{sec:methodology}

\draft{5}{Our research question is: \textbf{How do senior software engineering students experience working with AI on software architecture projects?} To answer it,} we studied student development teams in a third-year undergraduate architecture course at \university, in which students were encouraged to use AI \draft{5}{while they designed and implemented} a modern full-stack web application. We position this work as an \textbf{exploratory case study} that uses surveys and focus groups to describe how student teams experienced the project. In line with case study guidelines~\cite{Wohlin:2022}, we first describe the course, then the instruments we used to collect data, then the reflexive thematic analysis we applied to it, and finally how we established the trustworthiness of that analysis.

\subsection{Context (The Course)}
\label{sec:course-context}

A case study requires a \emph{real-life context} in which data collection took place~\cite{Runeson:2012}, and we describe ours in detail because a detailed description of the context establishes the credibility of the findings and lets readers judge how far they transfer to their own setting~\cite{Nowell:2017}. Our context was a semester-long, third-year undergraduate course in the software engineering program of \university, taught in the fall semester of the academic year '25--'26 and taken by students from both the software engineering and the computer science program. The learning goals of the course include basic software design and architecture skills.

The course combined lectures, lab exercises, and a team project, and the project is what this study focuses on. Teams of four to five students had to design and implement a software system from the ground up, applying modern web-development principles and technologies, starting from a list of high-level requirements the teaching team provided. The teaching team assigned students to teams using the same start-of-course survey we use as a self-efficacy instrument (\Cref{tab:self-efficacy-survey-description}), ensuring diversity in programming capabilities and balancing teams by gender; students could nominate one preferred teammate from within their own lab section.

The project ran in two phases with two deadlines each. In Phase~I, teams first delivered a set of design documents describing a software design that satisfied the project requirements, and then a fully functional implementation of that design. In Phase~II, the teaching team added requirements, and teams delivered an updated design and an updated implementation. Phase~II was not a repeat of Phase~I, however. Teams primarily refactored and reworked the code they wrote in Phase~I: they had to identify at least three technical debt items, record them in their backlog, and address them. They extended their architecture documentation with a module view and a component-and-connector view, each accompanied by an element catalog, interface descriptions, rationale, and a behavior sketch, and they revised or deprecated the architecture decision records they had written in Phase~I. Two new capabilities were required as well: a Model Context Protocol server through which an AI agent could book rooms, and an external API endpoint. The two phases also ran on different schedules: Phase~I allocated two weeks each to design and implementation, whereas Phase~II allocated one week to design and three weeks to implementation.

For this offering, the course was redesigned to account for the widespread adoption of generative AI in industry~\cite{Miller:2025,Ferino:2025:STGT}. Its learning objectives were expanded to include developing software with generative AI tools, and students were encouraged from the start of the project to experiment with generative AI, especially for coding and implementation. The teaching team expected that this would let students spend more of their time on the system's high-level design and specification, instead of on the boilerplate coding that had required considerable effort from students in previous years.

When the course was redesigned, in early summer 2025, there were few or no established guidelines for incorporating generative AI into software design course projects like this one, so the teaching team had no materials from which to instruct students. \draft{4}{With no established practice to teach from, the teaching team instead encouraged students} to experiment with various tools for themselves, using student packs of popular generative AI coding tools (Copilot, Google Gemini, etc.). Halfway through the course, between the two phases, the teaching team delivered a short lecture and a lab covering emergent best practices for AI usage. Both were built around a single three-phase workflow (research, planning, and implementation): students first use AI to understand the problem, then have it write an implementation plan that the team reviews before any code exists, and only then let it generate code in small, individually verified increments. The lecture presented this workflow as the alternative to ``vibe coding'' and also covered context engineering, project-wide custom instruction files, retrieval-augmented generation, evaluating model output, and the Model Context Protocol. The lab applied the same workflow to a small web application, to shows students how to work with GitHub Copilot. The materials of these sessions are made available on Figshare.\footnote{\url{https://figshare.com/s/7f25856966525523ea37}}

\subsection{Data collection}

\begin{figure}[tb]                                                                                                                 
    \centering                                                                                                                       
    \includegraphics[width=\linewidth]{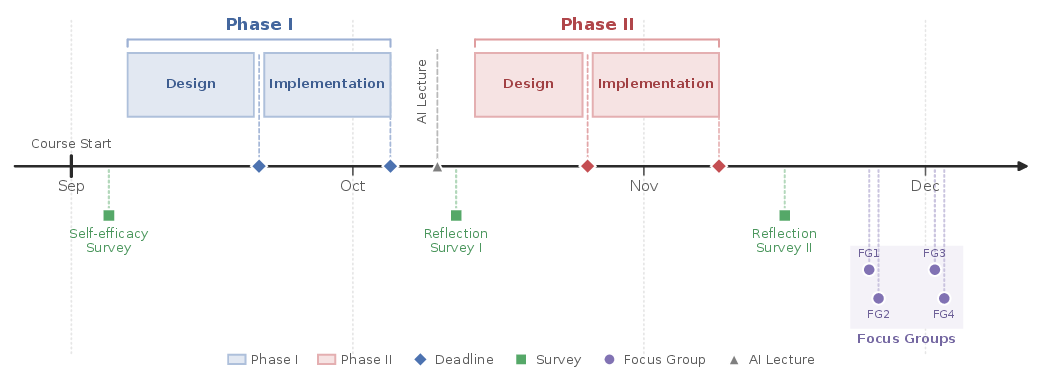}                                                                             
    \caption{Study timeline showing both the course deadlines and activities, and the data collection.}                                                
    \label{fig:timeline}                                    
  \end{figure}

We collected data using two methods at several points during the project, as \citet{Wohlin:2022} recommend, to build a corpus that describes students' experiences from more than one angle. Three surveys were administered before, during, and after the project, and after Phase~II we ran focus groups with the student teams to collect richer accounts of their experiences in an interactive group setting.  \Cref{fig:timeline} shows the course events (above the timeline) and the data collection activities (below it).

The ethical review board of \university\xspace approved the study.\footnote{Protocol number 25-0298} Participation was voluntary, and by consenting, students also allowed us to analyze the coursework they submitted. To ensure students did not feel coerced, a volunteer not involved in the course or the research distributed the consent forms, and the teaching team could not access the list of consenting students during the course.

\subsubsection{Surveys}

We administered three surveys: a \emph{self-efficacy survey} at the start of the course and a \emph{reflection questionnaire} after each project phase. The self-efficacy survey characterizes participants' at the start of the course, which helps interpret variation in how students adopted and used generative AI during the project~\cite{Runeson:2012}. The reflection questionnaires capture participants' self-reported AI usage and their perceptions of its impact after each phase.

\paragraph{Self-efficacy survey}

\begin{table}[t]
    \centering
    \begin{tabularx}{\linewidth}{p{3.5cm} X X}
        \toprule
        \textbf{Section} & \textbf{Format} & \textbf{Content} \\ \midrule

        \texttt{Prior Experience} & 3 questions with mixed formats: 2 frequency sliders (Never Used — Use Daily) and 1 four-option multiple choice & Asking participants to report their familiarity with version control tools and prior experience with web application development. \\

        \texttt{Programming Self-Efficacy} & 7 questions on a 7-point Likert scale (1 - not at all confident, 7 - very confident), adapted from \citet{Ramalingam:1998} & Asking participants to rate their confidence in a range of programming tasks, from writing a simple Python program to comprehending complex multi-module object-oriented code. \\

        \texttt{Software Design Self-Efficacy} & 9 questions on a 7-point Likert scale (1 - not at all confident, 7 - very confident), inspired by \citet{Bass:2021} & Asking participants to rate their confidence in software design tasks, from describing design patterns and architectural views to creating UML diagrams and designing an architecture from user stories. \\

        \texttt{Prior AI Usage} & 3 closed multi-select questions, with an optional other field & Asking participants whether they previously used generative AI, and if so, which tools and for which software engineering tasks. \\

        \texttt{Generative AI Self-Efficacy} & 5 questions on a 7-point Likert scale (1 - not at all confident, 7 - very confident), selected from \citet{Wang:2024} & Asking participants to rate their confidence in AI-assisted programming tasks, from fixing bugs with AI support to generating entire functional programs. \\

        \texttt{AI Attitude} & 3 questions on a 4-point Likert scale (1 - strongly disagree, 4 - strongly agree), inspired by \citet{Miller:2025} & Asking participants to rate their agreement with statements on AI usage orientations, from persistence when AI fails to verifying and adjusting AI outputs. \\

        \texttt{Industry Confidence} & 3 questions on a 7-point Likert scale (1 - not at all confident, 7 - very confident) & Asking participants to rate their confidence in obtaining a junior software engineering position, contributing to a development team, and using generative AI in a professional role. \\

        \bottomrule
    \end{tabularx}
    \caption{The sections present in the self-efficacy instrument, together with a description of the questions included in each section.}
    \label{tab:self-efficacy-survey-description}
\end{table}

\Cref{tab:self-efficacy-survey-description} summarizes the sections of the self-efficacy survey; the full verbatim instrument is in the replication package.\footnote{\url{https://figshare.com/s/7f25856966525523ea37}} The survey captures two types of pre-existing participant characteristics: factual prior experience and self-efficacy. \texttt{Prior Experience} covers distributed version control and modern web technologies, both of which the course required, and \texttt{Prior AI Usage} covers earlier use of AI for software development. Self-efficacy, the belief in one's own capability to execute a task~\cite{Bandura:1997}, is an important predictor of learning behavior and performance, and the three self-efficacy sections cover the core competency domains of the course: the assumed prior knowledge (programming), the learning objectives (software design), and the phenomenon under study (AI usage). To ensure construct validity, their items were adapted from \citet{Ramalingam:1998} for programming, inspired by the knowledge and skill framework of \citet{Bass:2021} for software design, and selected from \citet{Wang:2024} for generative AI.

\paragraph{Reflection questionnaires}

\begin{table}[t]
    \centering
    \begin{tabularx}{\linewidth}{p{3.5cm} X X}
        \toprule
        \textbf{Section} & \textbf{Format} & \textbf{Content} \\ \midrule

        \texttt{AI Usage} & 3--4 questions with mixed formats: 1 binary (Yes/No) and 2--3 open-text fields & Asking participants whether they used generative AI during the phase, which tools they used, for which tasks, and (Phase~I only) how they would describe the overall role of generative AI within their team. \\

        \texttt{AI Outcomes} & 5 statements on a 5-point Likert scale (Strongly Disagree --- Strongly Agree) & Asking participants to evaluate five outcomes of their AI use: time savings, support for learning about software architecture, contribution to team goals, the overhead of verifying and correcting AI outputs, and the degree to which AI influenced architectural decisions. \\

        \texttt{Open Reflection} & 2--3 open-text questions (partially conditional on AI use) & \draft{1}{Asking participants, in Phase~I, who found AI useful to describe why and which strategies they applied, and who did not find it useful to explain why and what they would change to make it more useful; and, in Phase~II, to describe the challenges they encountered and how they addressed them. Participants who did not use AI are asked to explain why not.} \\

        \midrule
        \multicolumn{3}{c}{\textit{Phase~II only}} \\
        \midrule

        \texttt{Longitudinal Change} & 3 questions with mixed formats: 1 binary (Yes/No) and 2 open-text fields (one conditional) & Asking participants whether their AI usage changed between Phase~I and Phase~II, and if so, how (e.g., changes in reliance or prompting strategy). Designed to capture the influence of the AI-skills lecture delivered between phases on students' usage. \\

        \bottomrule
    \end{tabularx}
    \caption{The sections present in the reflection questionnaires, together with a description of the questions included in each section. The bottom section is exclusive to the Phase~II questionnaire.}
    \label{tab:reflection-survey-description}
\end{table}

\Cref{tab:reflection-survey-description} summarizes the sections of the reflection questionnaires; the full verbatim instruments are in the replication package. \texttt{AI Usage} records whether, with which tools, and for which tasks students used generative AI in each phase. \texttt{AI Outcomes} asks students to rate the impact of their AI use on productivity, learning, and architectural decisions, and \texttt{Open Reflection} lets them describe strategies and challenges that Likert items alone cannot capture. The Phase~II questionnaire adds \texttt{Longitudinal Change}, which tracks whether and how students' AI usage changed between phases, and so whether the lecture delivered between them affected their practice.

\subsubsection{Focus Groups}

To complement the survey data with richer, qualitative accounts, we ran four focus groups with eight student teams after Phase~II. Focus groups suited the group-based nature of the project. Compared to individual interviews, they let participants build on and challenge each other's accounts, surface a team's consensus and disagreements, and reach a shared sense-making that individual interviews cannot access~\cite{Stewart:2015}.

\paragraph{Team recruitment}

\draft{4}{The teaching team announced the focus groups to the whole class and explained their purpose. Participating teams received cookies and soda, and their focus group replaced a group reflection assignment.} Of the 16 teams, 11 expressed interest, and the teaching team helped select eight of them. Because focus groups are group-based, we did not want them to re-trigger interpersonal conflicts or place undue stress on participants. \draft{5}{Therefore, one reason not to select all teams is to} exclude teams that had experienced interpersonal conflict during the project or had members who did not contribute. This might bias the focus groups toward more cohesive teams, a risk we accepted to avoid confrontational or uncomfortable sessions for the participants.

The eight teams were distributed across four sessions, two teams per session, yielding 8 to 10 participants per session. This kept sessions within the recommended range of 8 to 12 participants~\cite{Stewart:2015}: below it, one or two voices tend to dominate, and above it, active participation by all members becomes difficult to manage. Four sessions across eight of the 16 teams also gave a wide variety of perspectives on the project~\cite{Stewart:2015}.

\paragraph{Script Design}

The focus group script follows the guidelines of \citet{Braun:2022} and \citet{Stewart:2015} and has five sections (\Cref{tab:focus-group-script}). It orders questions from general to specific, as \citet{Stewart:2015} recommend, and moves participants from concrete, shared recollection toward deeper individual reflection. Each session opened with an anchoring exercise in which participants recalled an architectural decision they had made, which was meant to minimize retrospective bias~\cite{Flick:2000}, and then moved to a group discussion on team dynamics, AI use, and motivation. Each session closed with a synthesis exercise in which participants wrote down one or two words, or a metaphor, characterizing the role generative AI had played for them during the project, and then shared and responded to each other's characterizations. We refer to these as participants' \emph{closing metaphors} where we draw on them in \Cref{sec:findings}.

Because each session combined two established teams, participants knew their own teammates but not the other team, which risks within-team clustering during the discussion~\cite{Stewart:2015}. To counteract this, discussion prompts explicitly invited participants from different teams to react to and build on each other's accounts, and follow-up probes were directed across teams to surface differences and similarities in how the two teams approached the project. This cross-team dynamic also invited disagreement about experiences the two teams did not share.

The script was piloted once with junior researchers, which led to minor adjustments so that all five sections fit within the allotted 90 minutes. \citet{Stewart:2015} recommend piloting with respondents representative of the target population, but we had limited access to students outside the study population.

\begin{table}[t]
    \centering
    \begin{tabularx}{\linewidth}{p{3.5cm} X X}
        \toprule
        \textbf{Section} & \textbf{Format} & \textbf{Content} \\ \midrule

        \texttt{Anchoring Exercise} & Individual recollection of an architectural decision, followed by a round-robin sharing. & Participants individually jot down one key architectural decision or feature from their project, and then go round and share their decisions. Sharing grounds the subsequent group discussion in specific, concrete recalled experiences rather than abstract perceptions of the project and AI. \\

        \texttt{Unpacking Decisions} & Structured group discussion centered around the previously shared decisions, with moderator probes directed across both teams. & Explores team dynamics and AI's role in decision-making from two angles. The first covers decisions that generated notable disagreement (broadly interpreted as difficult, time-consuming, or revisited), probing team process, shared AI norms, and whether AI contributed to or complicated conflict. The second covers decisions for which the team relied heavily on AI, probing how AI was used for architectural reasoning, how outputs were accepted or rejected, and whether AI use for design differed from use for coding. Probes actively invited the other team to compare or contrast their own experience. \\

        \texttt{Evolution of AI Use} & Group discussion with moderator probes & Participants describe changes in how they used generative AI between Phase~I and Phase~II: what prompted those changes, whether outputs improved, and any adjustments to prompting strategies or task allocation. \\

        \texttt{Motivation} & Group discussion with moderator probes & Participants reflect on whether they genuinely cared about the quality of their work and what drove or undermined their investment in the project, providing motivational context for AI adoption choices. \\

        \texttt{Synthesis Exercise} & Individual reflection followed by brief group sharing, and a final open question & Each participant writes one or two words, or a metaphor, characterizing the role AI played in their project, shares with the group, and briefly responds to each other's characterizations. A final open question invites participants to raise anything not covered in the discussion. \\

        \bottomrule
    \end{tabularx}
    \caption{The sections of the focus group script, together with a description of their format and content. The script was designed to move participants from concrete recollection (anchoring exercise) toward progressively deeper reflection on decisions, longitudinal change, and motivation, exploring the experience of AI usage from three complementary angles: teamwork dynamics, decision-making, and motivation.}
    \label{tab:focus-group-script}
\end{table}

\paragraph{Procedure}

Participants gave written informed consent before each session. The moderator then explained the purpose and scope of the session, invited participants to respectfully disagree with each other, and reiterated that there were no wrong opinions. Sessions lasted 90 minutes and took place in a large, well-lit lab room where participants and the moderator sat around a square table. To encourage interaction across the two teams, participants were assigned seats alternating by team.

Each session was run by a moderator and a facilitator~\cite{Stewart:2015}. The moderator led the discussion, kept it on track, ensured all participants were called on, and managed the time. The facilitator sat to the side of the room, took notes, and was occasionally called on to ask summarizing questions. All sessions were audio- and video-recorded; the video captures non-verbal cues, such as gestures and reactions, that supplement verbal responses~\cite{Stewart:2015}. These recordings were the primary data source for the analysis, and \Cref{sec:data-analysis} describes their transcription.

\paragraph{Researcher Positioning}

In line with \citet{Braun:2022} and \citet{Stewart:2015}, we describe the positions and backgrounds of the moderator and facilitator. The first author moderated all sessions and later coded the data; they are a software engineering researcher with extensive experience teaching software engineering courses and working with software architecture in industry. This background helped them interact with participants, for instance by letting students know they could use domain-specific terminology, and it informed their coding of the transcripts (\Cref{sec:data-analysis}). The second author facilitated the sessions and is an experienced qualitative researcher. Neither was involved in the course, so there was no prior relationship between them and the participants.

\subsection{Data Analysis}
\label{sec:data-analysis}

We analyzed the data along two strands. The closed-response survey items were analyzed descriptively, to characterize the participants and the context they worked in\draft{5}{. The} qualitative data (\draft{5}{the focus group transcripts and the responses to the open survey questions}) were analyzed through reflexive thematic analysis, to construct an account of what the project was like for participants.

\subsubsection{Descriptive Analysis of Closed-Response Data}

The closed-response items of the three surveys (the Likert-scale ratings of programming, software design, and generative AI self-efficacy, the AI outcome perceptions, and the binary and multi-select items on prior and in-course AI usage) are summarized with descriptive statistics. We report them after the themes, in \Cref{sec:results:descriptive}, to situate participants' accounts in relation to their self-efficacy and self-reported AI usage.

\subsubsection{Thematic Analysis of Qualitative Data}

We analyzed the open survey questions and the focus group transcripts using reflexive thematic analysis, following the framework of Braun and Clarke~\cite{Braun:2013, Braun:2022} as applied in comparable software engineering education studies~\cite{Ahmed:2025, Pearson:2025}. We believe its inductive and reflexive approach suits our interest in the lived experiences of students. Because the study is exploratory, thematic analysis lets us identify emerging patterns of meaning in what students share, and it lets us draw on our experience as software engineering researchers to interpret their accounts.

\paragraph{Analytic Orientation and Reflexivity}

Our coding and theme development were primarily inductive: given the exploratory nature of the study, we started from participants' accounts rather than a pre-selected theory or framework, and we constructed the two themes from the patterns of meaning we saw across those accounts. We selected self-determination theory~\cite{Ryan:2000} only when we noticed that some statements related to its constructs: that participants had little say over the pace of the project, that they did not feel able to do the work the project asked of them, and that they had stopped caring about what they were building. \citet{Braun:2022} note that a theoretical lens need not be settled in advance and may be adopted during interpretation where it affords a richer account. Our analysis is therefore inductive in its coding and theme development and theoretically informed only later: we report the themes in \Cref{sec:findings} in participants' own terms and use self-determination theory in \Cref{sec:discussion} as a lens to interpret them.

Because the setting is educational, we take a more contextualist view. To interpret how students experienced the course, we relied on information from the teaching team about what participants went through and how they constructed working with AI. The first author, who coded the data, also drew on their own teaching and industry experience (see Researcher Positioning above) while coding the transcripts.

\paragraph{Pre-analysis: Transcription and Preparation}

The audio and video recordings of the focus groups were transcribed with Whisper~\cite{Radford:2023}, run locally because the recordings contain personally identifiable information and the consent forms did not allow the recordings or transcripts to be shared with external services. The first author then corrected the automated transcripts by hand and annotated them with contextual information visible only in the video, such as participants nodding in agreement, raising their hands, or reacting non-verbally to others' statements. In a second pass, all data collected for this study (surveys and focus groups) were pseudonymized: each participant received a unique identifier that replaced their name and identifies them across the surveys and focus groups.

\paragraph{The six stages of Braun and Clarke}

\citet{Braun:2022} identify six phases of reflexive thematic analysis. The phases are described in order but are not linear; as the analysis developed, we moved between them several times to refine codes and to find and name themes. Below we describe what we did in each phase.

\begin{itemize}
    \item \textbf{Phase 1: Dataset familiarization.} The first author read all survey responses and transcripts and rewatched each focus group recording. This happened a few weeks after the focus groups were held and transcribed, and so served as an effective reminder of the sessions. During this process, the first author wrote free-form memos recording initial observations, impressions, and emerging questions about the data. Prolonged engagement with the data, and documenting reflective thoughts about potential codes while becoming familiar with it, are the means \citet{Nowell:2017} describe for establishing trustworthiness in this phase. The memos already contain observations that became part of the final two themes, for instance the notion that students felt gaslighted by AI: \emph{AI will suggest changes to everything. As one participant says: ``AI gaslights you''}.

    \item \textbf{Phase 2: Data coding (two passes).} The first author coded the data in two passes. The first pass produced descriptive (semantic) codes that stayed close to the data, and a selection of relevant excerpts from the transcripts. The first and second author then discussed the more interesting codes and excerpts. Because our coding is reflexive, the goal of this discussion was not to measure agreement~\cite{Braun:2022} but to consider alternative explanations and deepen our understanding of what participants were saying; discussions of this kind are the peer debriefing \citet{Nowell:2017} recommend throughout coding. The second pass served as a completeness check and an opportunity to extract latent meaning not apparent in the first pass, for instance by coding more of the emotions participants expressed and by identifying every passage in which they talked about AI reliance. Working through the dataset twice also follows \citet{Nowell:2017}, who recommend that coding be scrutinized at least twice and that every data item receive full and equal attention before themes are treated as settled. The second pass resulted in 1,775 coded excerpts with 1,501 unique codes.

    \item \textbf{Phase 3: Initial theme generation.} The first author grouped the codes into seven candidate themes, each a recurring pattern across the data that we believed relevant to the research question, such as how working with AI made participants feel, whether they controlled AI, and how the course structure influenced their AI usage. The first two authors discussed these candidate themes and their excerpts repeatedly, to work out how each should be defined and what united the codes within it. They were not themes yet, but they captured experiences participants repeatedly shared and were instrumental in constructing the final two themes.

    \item \textbf{Phase 4: Theme development and review.} To assess whether the candidate themes held up against the full dataset, the first author created Miro boards for the seven candidate themes, grouping similar codes, annotating the groups with interpretations, and visualizing the relations between them. \citet{Nowell:2017} recommend diagrams of this kind to make sense of the connections between codes and themes and to keep a record of how concepts and themes developed. The boards also made clear which focus groups or participants each code came from, and it was through them, and through a final board, that a structure of two contrasting themes centered on AI reliance was constructed. Each board was discussed extensively with the second author to verify that the underlying data supported the interpretation, and in later iterations with all authors to agree on a shared interpretation of the data. Returning to the raw data in this way is the test of referential adequacy \citet{Nowell:2017} describe for this phase, and vetting the boards with the other authors follows their recommendation that themes be reviewed by the wider team.

    \item \textbf{Phase 5: Theme refining, defining, and naming.} Starting from the Miro boards, the first author wrote several draft summaries of the two emerging themes. Through these drafts, together with the visual summaries, the definitions of the final two themes emerged, and we settled on the insights we combine to form each theme. Keeping the drafts documents how the themes were defined and named, which \citet{Nowell:2017} list among the means of establishing trustworthiness in this phase.

    \item \textbf{Phase 6: Writing up.} For each theme, we selected representative quotes and combined them into a narrative that conveys both the theme's meaning and its significance for understanding students' experiences with generative AI during the project.
\end{itemize}

\paragraph{On the usage of generative AI for reflexive thematic analysis}
A recent mapping study shows that generative AI is being applied to thematic analysis in software engineering~\cite{Barros:2025}, and custom-made tools exist to support this~\cite{Sharma:2025}. However, \citet{Montes:2025} found several limitations: generative AI is not suitable for latent interpretation, and when it constructs themes, there are no clear boundaries between them.  \citet{Ernst:2026}  likewise conclude that generative AI lacks the analytical depth reflexive thematic analysis requires. Braun and Clarke themselves reject its use for reflexive thematic analysis, arguing that the meaning-making the method requires is not something generative AI is capable of~\cite{Jowsey:2025}. In light of these findings and critique, we did not use generative AI for any of the coding or construction of the themes. In fact, we only used AI to evaluate and critique early drafts of our prose, including the descriptions of the themes, to help improve the flow and clarity of our writing.

\subsection{\draft{2}{Trustworthiness of this study}}
\label{sec:trustworthiness}

Case study research in software engineering is commonly evaluated against construct validity, internal validity, external validity, and reliability, or, for flexible design studies, against credibility, transferability, dependability, and confirmability; \citet{Runeson:2012} treat the choice between the two as one of terminology. Both frameworks assume that a stronger analysis is one that depends less on the researcher who produced it. Reflexive thematic analysis makes the opposite assumption: the researcher's subjectivity is a resource for the analysis, and a second analyst working the same data would produce a different account, and not necessarily a better or worse one~\cite{Braun:2022}. For this reason we do not compute inter-rater reliability, as there was only one coder, and we make no claim of saturation, a concept \citet{Braun:2022} consider incompatible with reflexive thematic analysis and one we never tried to reach.

Instead, we evaluate this study against the criteria \citet{Braun:2022} define for reflexive thematic analysis, and we follow the practical guidelines of \citet{Nowell:2017}. From \citet{Nowell:2017} we take the phase-by-phase means of establishing trustworthiness, and \Cref{sec:data-analysis} reports where each was applied: prolonged engagement with the data, triangulation of the data collection modes, memo writing during familiarization, peer debriefing between the first two authors, diagramming how codes and themes connect, a return to the raw data to test referential adequacy, and documentation of how the themes were defined and named. The criteria of \citet{Braun:2022} ask that transcription be thorough and checked against the recordings, and that coding give repeated attention to every data item rather than build themes from a few vivid examples (\Cref{sec:data-analysis}); that themes cohere around a central organizing concept, that data be interpreted rather than summarized, and that the extracts evidence the claims made (\Cref{sec:findings}); and that the approach and its theoretical assumptions be stated explicitly, that what was claimed match what was done, and that the researcher be positioned as active in the analysis (\Cref{sec:limitations}).

\Cref{sec:findings} reports the findings, first the two themes constructed through reflexive thematic analysis and then the descriptive statistics.

%% file: sections/5-results.tex
\section{Results}
\label{sec:findings}

\draft{4}{\draft{5}{From the focus groups and the open survey responses}, we constructed two themes about how students experienced working with AI, both centered on participants' judgment over what AI produced. We first give an overview of both themes and each theme's \textbf{central organizing concept}: the core idea that unites a theme as a pattern of shared meaning~\cite{Braun:2022}.}

\draft{4}{The first theme, \textbf{Surrendering Judgment to AI}, is organized around \draft{5}{the accounts of participants who} spoke about their extensive and deepening reliance on AI, which we interpret as participants eventually surrendering their judgment to it. Participants who said they felt unprepared for the project described ceding decision-making to AI, no longer verifying its output, and eventually becoming stuck on bugs in a codebase they did not understand. To get unstuck, they often relied on AI even more instead of repairing their understanding, so the surrender only deepened their reliance. We believe an important element of this theme is how participants spoke about their work after that surrender: they described feeling ashamed, frustrated, and emotionally disconnected from work that had started to feel pointless and empty. In our reading, this emotional toll left them with even less reason to break the cycle of AI reliance.}

\draft{3}{In contrast to this surrender, the second theme, \textbf{Applying Human Judgment}, is organized around \draft{5}{the accounts of participants who never surrendered their judgment to AI, and of participants who reclaimed it from AI.} The core observation in this theme is how this judgment was anchored in the knowledge and experience of the participants. We observe how some participants refer to knowledge they brought into the course, while others said they recognized during the project that they lacked knowledge, and acquired it. In this theme, participants shared how they felt changes to the course gave them the space to acquire knowledge, and how their judgment afforded a different way of working with AI. One in which they did not surrender their judgment to it. }

Because we constructed the themes inductively, surrender as a construct was not discussed with participants during the focus groups. Therefore, accurately counting the number of participants who surrendered during the course, or grouping participants into groups based on whether they surrendered, is therefore not possible, nor was that our goal. Reflexive thematic analysis discourages counting how often codes occur~\cite{Braun:2022}, which is another reason why we do not definitely attempt to assign participants to either theme. The transcripts do show that participants' experiences varied: some surrendered their judgment and later reclaimed it, some never surrendered it, and some surrendered it and never reclaimed it. Several participants, among them P5, P6, P9, P19, P32, P37, P44, and P45, speak in both themes. P6, for instance, described their reliance on AI as a ``slippery slope'' in the first theme, and in the second theme described ``wanting to write more of the code by my own hands''.

\begin{figure}[tp]
    \centering
    \includegraphics[width=\linewidth]{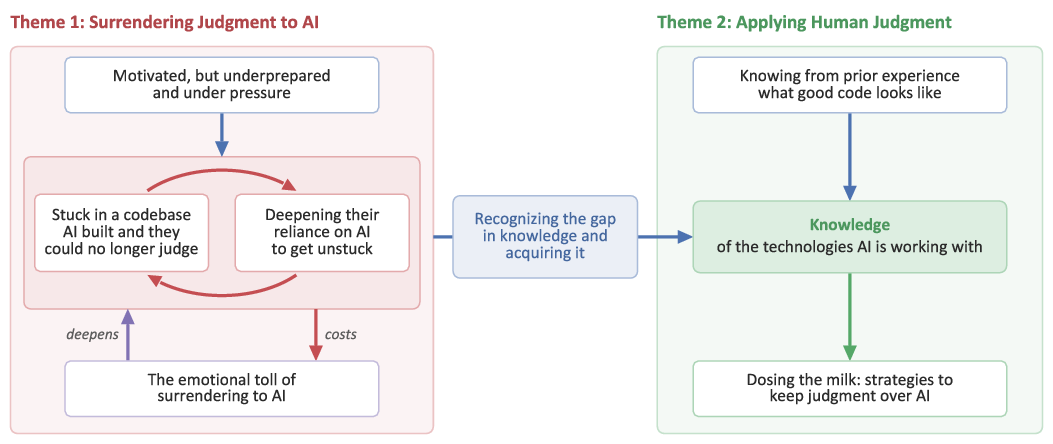}
    \caption{Overview of the two themes and their insights. Theme 1 is \draft{4}{a cycle}: participants who arrived motivated but underprepared and under pressure surrendered their judgment to AI and became stuck in a codebase they could no longer judge, and what they described feeling after that surrender left them, in our reading, with even less reason to break the cycle. Theme 2 builds on knowledge: judgment over AI rests on knowing the technologies AI works with, and participants discussed the strategies that helped them apply it. The insight that connects the two themes, \draft{5}{recognizing the gap in knowledge and acquiring it, is how some participants said they broke the cycle; others brought that knowledge} into the course with them.}
    \label{fig:theme-overview}
\end{figure}

\draft{3}{Figure~\ref{fig:theme-overview} shows both themes, the insights that make up each, and the insight that connects them. In the remainder of this section, we quote the participants to explain how each theme is constructed.}

\subsection{Theme 1: Surrendering Judgment to AI --- a deepening reliance on AI that sustained itself}

\draft{3}{This theme is constructed around three insights that follow participants from the conditions at the start of the project, to being stuck in a codebase they could no longer judge, to \draft{4}{the emotional toll they described after that surrender, which we read as leaving them with even less reason to break the cycle}.}

\subsubsection{Motivated, but underprepared and under pressure}

\draft{1}{In these accounts, participants describe three conditions that we believe are relevant for understanding what later led them to surrender to AI: the motivation they started with, the gap they felt between their own skills and what the project asked of them, and the pressure of the deadlines.}

\begin{pquote}{P18 -- Focus Group}
     I think yeah I was excited to learn new things. That was my motivation to try out uh like typescript [at] first like I wanted to learn uh because nowadays typescript is more widely used so i was willing to give [it] a chance and learn as much as i could over the span of this project.
\end{pquote}

\noindent Referring to the start of the project, some participants described being motivated to learn new skills. For P18, the project was a chance to learn skills they considered relevant to their future work. Moreover, we think that the recent widespread adoption of AI by industry and expectations that developers can use AI for software development will have contributed to this motivation, as some participants state that they were motivated because they were allowed and encouraged to use AI. \iquote{P38 -- Focus Group}{But also, I was just going to say that being encouraged to use AI made me more motivated because it was the first course that I was allowed to use AI.}
\draft{1}{Both quotes point to participants who initially felt motivated to work on the project. For different reasons, we read each as describing an interest in the work itself.}

While talking about the start of the project, participants also mentioned their experience and confidence. To enroll in the course, participants were required to have passed prior programming courses and, in the first week of the course, almost all participants reported confidence in their own programming skills (\Cref{fig:self-efficacy-scores}). When designing the course, the teaching team expected these prior programming skills to transfer to the web development skills required for this project.
However, despite this experience and initial confidence, participants reported not knowing how to start their projects.
\iquote{P9 -- Focus Group}{Some of us have never even started like a 
JavaScript like project or something like that. So just like getting that started is even just a big barrier.}
For this participant to refer to starting as a ``big barrier'' suggests that, initially, they felt their existing programming knowledge did not transfer to web development, and that the project sat outside their comfort zone. We read the gap between what participants brought and what they felt the project asked of them as a sign that they did not feel confident in their own abilities, nor able to do the work the project asked of them. It also hints at a mismatch between what the course team expected of participants and what participants felt they could do. P5, who was themselves already experienced with this stack (as we return to in the second theme), observed how this shortfall was shared across the class, and how important AI became:

\begin{pquote}{P5 -- Focus Group}
we didn't know anything about the software architecture a lot of the people in the class didn't have any experience with like TypeScript or Docker or like any of the technology we were using [...] and then AI was like the only glue that would like somehow progress forward
\end{pquote}

\noindent We read P5's observation as showing how a shortfall of experience translated directly into an early reliance on AI. With unfamiliarity in TypeScript and Docker, and the software architecture itself something the class did not yet know, AI became the ``glue'' holding the work of the team together, the single thing that let the class ``somehow progress forward''.

Finally, we think the course structure, especially with respect to the deadlines, further contributed to how participants felt at the start of the project.
\iquote{P37 -- Focus Group}{I felt like we had a very short amount of time to make a lot of really big decisions about the system and then had to jump right into implementation.} It is this feeling of having limited time that we think may also have left participants with little say over how they worked: describing ``a very short amount of time'' for ``really big decisions'', and then having to ``jump right into implementation'', P37 hints at a pace that was set for them rather than by them. Combined with the accounts of missing experience, this makes us believe participants experienced considerable stress at the start of the project. \iquote{P9 -- Focus Group}{I feel like in the context of the course, it was like a bit, I feel like a word came to mind like a bit overwhelming.} P9 goes as far as to describe the start of the course as ``overwhelming'', which reinforces our sense that at least some participants felt they had neither the experience the project required nor much say over how they worked on it.

The course was designed with a clear idea around AI usage: participants could use AI for the implementation, especially to work with a new programming language, while they were encouraged to use the course material for the software architecture. For some participants, this was part of what made the project motivating at the start. However, the way these participants describe starting the project suggests that this was not a boundary they found easy to hold. Coupled with how participants experienced the start of the course, with them describing not feeling competent in the technologies the project required, and not feeling much control over its pace, we think the motivation participants started with was already under strain, a shift we return to in the third insight. 

\subsubsection{\draft{5}{A reliance on AI that deepened itself}}
\draft{1}{Based on how participants described using AI once they were working on the project, we construct an account of how their initial reliance on it pulled them into a self-reinforcing cycle.}

\begin{pquote}{P36 -- Focus Group}
    I used it[AI] quite extensively at the start of the project, trying to get all the different tools configured and work together to get a baseline project.
\end{pquote}

\noindent Given how some of the participants experienced the start of the project, unfamiliar with the stack, and under deadlines that left them little room, any tool that gets a baseline project running is valuable, and that is what P36 describes using AI for. \iquote{FG47 -- Focus Group}{I think my biggest reason for preferring to go to AI over, let's say, asking a teammate or a TA or prof is it's just, well, there's a huge time constraint. There's due dates and we all felt that sometimes it was too short of a due date or too short of a time working on the project.} This quote from FG47\footnote{\draft{5}{Identifiers starting with FG denote participants who consented to take part in the focus groups, but not to the use of their survey data.}} illustrates how participants experienced the conditions at the start of the course as a reason to use AI. Crucially, though, while this initial turn to AI might have helped some of the participants get started, for some it also became a deepening cycle of reliance on AI. Across the various groups, participants described becoming reliant on a tool they found useful and powerful, but at a cost. \draft{1}{The closing metaphor from P22 (\Cref{tab:focus-group-script}) names the ambiguous role AI came to occupy:} 

\begin{pquote}{P22 -- Focus Group}
    Um, I said I felt completely dependent on it[AI], and that it was my best friend and greatest enemy.
\end{pquote}

\noindent To us, holding ``best friend'' and ``greatest enemy'' together in a single sentence, alongside feeling ``completely dependent'', names the bind that runs through this insight: the tool participants could not work without was also the one causing them trouble, and being dependent on it meant they could not simply put it down.

To understand what this reliance was like, we begin with how aware participants were of it. Probes in the focus group asked participants to describe how they used AI while making architectural decisions, and specifically whether they allowed AI to make those decisions for them. Across the focus groups, participants repeatedly denied allowing AI to make architectural decisions; P25, for instance, describes using AI as an information source rather than ceding decision-making to it.

\begin{pquote}{P25 -- Focus Group}
    So I think in that way, we kind of used AI to do some research on some differences we might experience based on our choice.
But I think overall, the actual decisions were more based on our own knowledge of them.
\end{pquote}

\noindent However, when we prompted participants to give examples of their AI usage, those examples complicate this account: \iquote{P34 -- Survey}{[I used AI] to find out what files are needed to get a working backend structure, and what to fill into those files}
The initial structure of a backend, in terms of files and modules, determines the architecture, especially for a TypeScript project. We do not read this as participants misrepresenting their own process; rather, under the pressure of an unfamiliar project and tight deadlines, relying on AI to scaffold the backend implicitly cedes these impactful architectural decisions to it, perhaps without participants recognizing how important those decisions were. In this case, the initial design of the backend also specifies the interface between the backend and the frontend, and therefore core components of the system's architecture.

As some of the participants worked with and relied on AI for their projects, they eventually found themselves unable to make sense of the AI-generated information.
\iquote{P21 -- Survey}{It[AI] was extremely helpful yet confusing cause I still didn't understand what we were exactly doing, it was just creating code for us!}
We read the combination of ``extremely helpful'' and ``still didn't understand'' in a single sentence as showing how far the reliance went. We do not read this as the participant being unaware of the gap, as they name it themselves, but as this participant pushing on without fully understanding what AI did: AI ``was just creating code for us'', and the work continued. This is where, in our reading, ceding decision-making shades into surrendering judgment, because code a participant cannot follow is also code they cannot evaluate. P21 is not the only participant who did not fully evaluate what AI produced: \iquote{P32 -- Focus Group}{I'd actually say that I probably don't register how often it[AI] is incorrect.}

Continuing while suspecting AI might be wrong was not something only P32 described. When we asked the focus groups what they did with information they could not check for themselves, this is what participants shared:

\begin{pquotes}
    \utt{Moderator}{If you prompted AI with a task description, and it gave you back some information that you
maybe didn't really know how to validate or how to qualify, what did you do in those cases?}
 \utt{P14}{I would try to keep going.}
 \utt{P22}{Cross your fingers}
 \utt{Comment}{P16, P11, P12 and P19 all nod in agreement. }
\end{pquotes}

\noindent \draft{5}{This interaction, and the nods that accompanied it, suggest that several participants in this session chose to continue working with information they were not able to fully process:} they were aware that AI might give them incorrect information, yet still accepted that possibility.

These quotes indicate that, at least initially, participants were able to get away with not verifying, or only very shallowly verifying, AI-generated information. However, this behavior eventually led to issues for participants. \iquote{FG49 -- Focus Group}{But if you tell the AI to do it, it'll just go. And then it'll complete it. And you don't actually see the issues with that design because it's just done it for you.} What FG49 describes is not a suspicion that something is wrong, but the absence of any occasion to notice: because AI would ``just go'' and complete the task, the problems it introduced into a design were never surfaced for the participant to see. Speaking about it afterwards, FG49 can say the issues were there.

Those problems still had to be dealt with eventually. However, given how some of these participants were using AI, and how little experience they had with the stack, we think they were working with a codebase they had produced but could not fully understand. \draft{4}{This is the tension in how participants used AI: the reliance that let them create the codebase at the start of the project was also what began to cause their problems.} Dealing with these problems was challenging: \iquote{P44 -- Focus Group}{A lot of like the debugging is like, oh, here's this error that I'm facing. And then you make a change, and then you get a new error. And I don't really know if I'm stepping further or closer to my end goal, I guess.} In this case, the participant describes being unable to make sense of the AI-generated information, but maybe more importantly, also not knowing whether relying on AI was helping them solve their problem. This situation, of being stuck without knowing whether they were making progress, was experienced by several participants during the project: \iquote{FG52 -- Focus Group}{[I would do] whatever docker would allow. Like it would give you an error we try and fix it and then it give a different error try and fix that}.

For some participants, working through errors with AI did not merely stall progress but escalated the problem itself, as FG50 recalls another Docker issue: \iquote{FG50 -- Focus Group}{we couldn't give all of the information so maybe AI got confused [...] so that's where it[AI] started hallucinating stuff and that like kind of led to a lot of bad answers through that and then it just kept spiraling and we ended up in like a much bigger problem than what we started with}.
Looking back, FG50 can name what went wrong, though even in hindsight the account stays hedged: it is ``maybe'' that AI got confused. What we take from this is that the recognition came late. By their own telling, the answers ``kept spiraling'' into ``a much bigger problem than what we started with'' before the hallucination was caught, which suggests there was no point during the incident where the participant could have caught it.

We read these accounts as showing what that reliance cost participants: having lost their grip on the code, they became even more dependent on AI to solve the bugs and issues they ran into \iquote{P22 -- Survey}{we relied heavily on AI to figure out how to do any of the development. No one in our group had development experience to the level that this project required so generative AI was a massive crutch for us.}
The phrase ``massive crutch'' names the dependence: a crutch is what you need in order to move at all, and P22 describes a group that could not have progressed, or resolved the bugs they hit, without it.

When participants were stuck and relying on AI like this, escaping rarely meant repairing their own understanding; instead, it often meant deepening their reliance on AI. When we asked participants whether they checked or validated the information AI gave them while stuck, 3FG50 described their first response: \iquote{FG50 -- Focus Group}{I think one of the first things that I did, because of the time constraints, was just use a different AI model. But it kept giving similar answers [as the first model].} This answer reveals how deep the reliance ran: prompted about verification, the participant describes consulting another AI model. Verification itself was delegated to AI, and, in this case, without payoff, as the models ``kept giving similar answers''. Even when switching models did resolve the problem, participants' own understanding was not repaired:

\begin{pquote}{P25 -- Focus Group}
    There is a problem I was having where I was trying to make a request to the database and it would fail because of the time was in the wrong format, and I could not figure out why it was in the wrong format [...] until I switched which agent I was using and then it immediately figured it out.
    The first model kept giving information about why it could be wrong and then implementing the fixes and it was just wrong every time [...] and then it just gave a very concise reason of why exactly it was wrong.
\end{pquote}

\draft{1}{These two accounts are why we consider the reliance self-reinforcing. Neither participant adjusted their AI usage or repaired their understanding when they got stuck; both reached further into AI. For P25 that worked, but the resolution stayed entirely within the AI: the new model gave ``a very concise reason of why exactly it was wrong'', while, by their own account, the participant had not been able to work out the problem themselves.} Other times, reaching further into AI reliance resolved nothing at all: \iquote{P26 -- Survey}{Once I started using it[AI] to program aspects of the project I had never done before (backend, database), it[AI] started changing a ton of things at once and it was difficult to know if it[AI] was on the right track. It[AI] never reached the desired functionality.} P26 describes the cycle running its full course without any payoff: the parts of the project they knew least about were the parts they handed over most completely, AI ``changing a ton of things at once'' left them less able to tell whether it was ``on the right track'', and at the end of it the functionality was still not there. Where P25 at least got a working answer they did not understand, P26 got neither.

\draft{4}{To us, these accounts also show the central organizing concept of this theme: \draft{5}{what participants surrendered to AI was their judgment. They stopped evaluating what AI produced, whether information or code, and once they had surrendered their judgment, their reliance on AI only deepened.}}

\subsubsection{\draft{5}{An emotional toll that left little reason to break the cycle}} \draft{1}{To understand what \draft{4}{surrender left participants with}, we must consider not only how they used AI, but also what they told us it felt like.} \draft{5}{After surrendering, participants described feeling ashamed that they relied on AI rather than on their own knowledge, and frustrated with an AI that was confidently incorrect, which some of them experienced as gaslighting. They also described an emotional disconnect from work that had come to feel pointless and empty. In our reading, this emotional toll left them with less reason to break their reliance on AI.}

Some of the participants appeared to be very aware of their own reliance on AI, and as P19 discusses, for them it triggered a sense of shame:  \iquote{P19 -- Focus Group}{From my perspective, shame I have to rely on it [AI] rather than my own knowledge.} For this participant, a novice developer, the issue is not that the tool was unhelpful but that leaning on it ``rather than my own knowledge'' was something to be ashamed of: what they say about AI is really a statement about themselves. P22 describes the same relation between AI's capabilities and participants' own knowledge, but at the level of a team: \iquote{P22 -- Focus Group}{None of us felt like we knew more than the AI did, so it was like, why would we not just trust it?} The trust here rests on a comparison, and it is a comparison P22 describes their whole team losing: no one in the team \emph{felt} they knew more. Together, the shame P19 names and the deference P22 describes show how some participants who were deeply reliant on AI felt they did not possess the knowledge required to do some of the work themselves.

Some of the participants who relied on AI also trusted it, and that trust ran into a well-documented property of LLMs: their tendency to be confidently incorrect~\cite{Zhou:2024}. \draft{1}{Based on some of the closing metaphors, we think these instances where AI was confidently incorrect were particularly challenging for some of the participants.} In fact, FG51 used the word ``gaslighter'' to describe AI:

\begin{pquote}{FG51 -- Focus Group}
{The term I started with [to describe AI] was like intelligent gaslighter because honestly it does feel like it then knows a couple of things it does give you it could be very helpful it gives you the answer you sort of want but it just sort of says like I'm right you're wrong trust me on this it should work the way I tell you to. }

\end{pquote}

\begin{figure}
    \centering
    \includegraphics[width=0.3\linewidth]{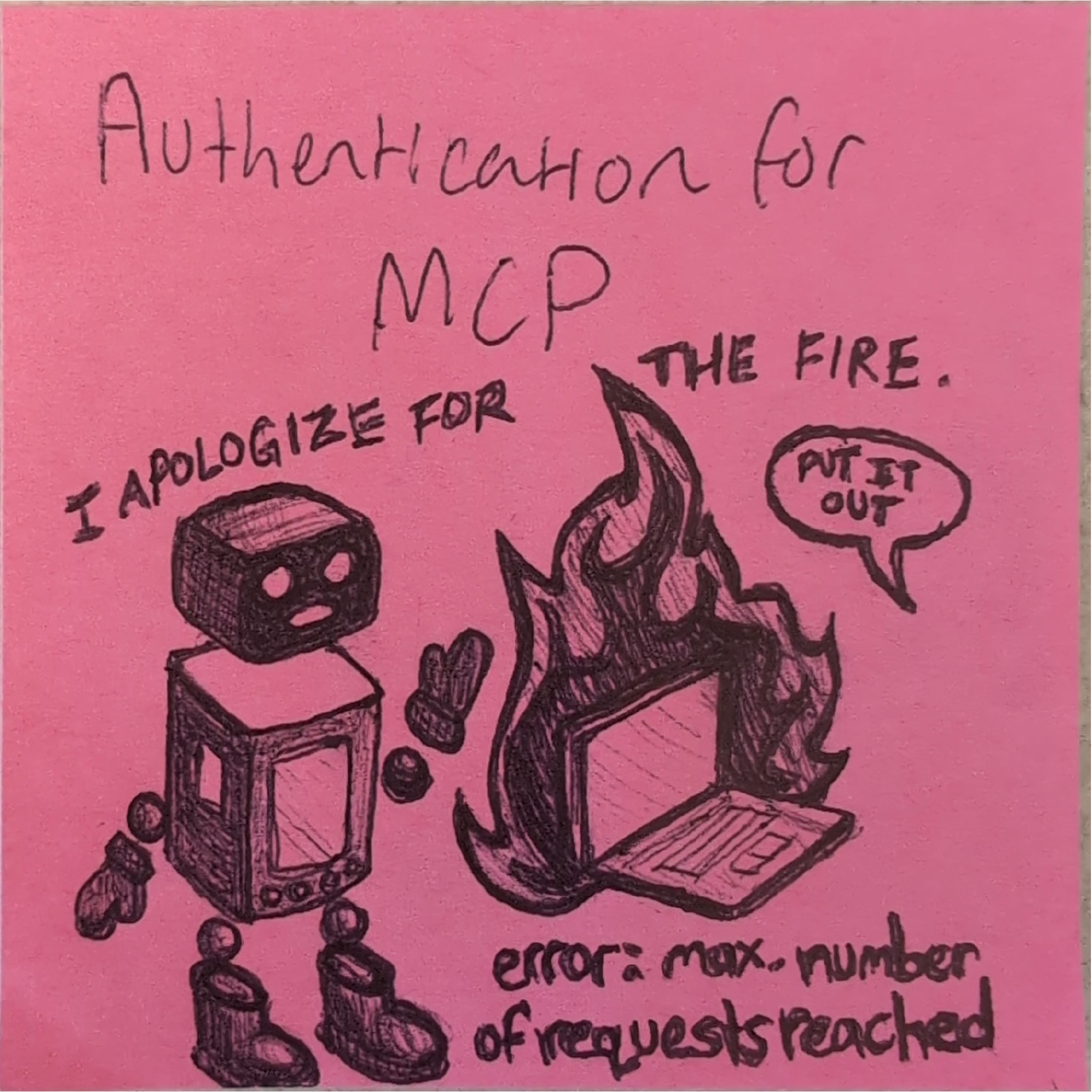}
    \caption{Participants of the focus groups were also given sticky notes to take notes; FG53 also drew the following picture in addition to sharing their metaphor. The top half of the sticky note contains some unrelated text related to a task this participant worked on during the project.  }
    \label{fig:burning-robot}
\end{figure}

\noindent We believe the notion of \emph{gaslighting}\footnote{Defined by Merriam-Webster: ``Gaslighting is a form of psychological and emotional abuse where a manipulator makes someone question their own memory, perception, or sanity. The goal is to establish power and control by forcing the victim to depend entirely on the abuser's version of reality.''} captures what FG51, and some of the other participants, were reaching for. FG53, who joined a different focus group, described AI in similar terms, and even drew a pictogram related to this idea (\Cref{fig:burning-robot}):

\begin{pquote}{FG53 -- Focus Group}
    I said Copilot is like that friend that you meet and think he's really smart. So you talk to him a lot, but then you get sick of him because you realize he doesn't actually know what he's talking about. And he's been manipulating you and spending a lot of time with you.
\end{pquote}

\noindent While AI does not have any malicious intent, which is why the definition does not fully apply, we do believe that it felt like gaslighting for some of the inexperienced participants when the AI they trusted was confidently incorrect and made them doubt their own knowledge. The following interaction with FG52 shows what that could look like in practice:

\begin{pquotes}
    \utt{FG52}{I think there were definitely times it[AI] wouldn't listen. It[AI] would say this[change to the backend] is necessary, which would make me consider it, but ultimately I wouldn't change the backend.}
    \utt{Moderator}{So when you say AI would suggest changing the backend, you said you would consider it. Why was this?}
    \utt{FG52}{Just because I didn't have perfect knowledge of what I was trying to do in the front end. You know, or what was actually needed. Obviously, I don't trust it[AI] 100\% but it has better knowledge than me, and it's telling me that I need to do something.}
\end{pquotes}

\noindent In this interaction, FG52 explains how AI kept insisting a change to the backend was needed, even though they thought it was not. Notably, this participant held their ground: ``ultimately I wouldn't change the backend''. What the interaction shows, then, is not that AI overrode their judgment, but what it cost them to keep it. Because they did not have ``perfect knowledge'' of what was needed, a tool they granted ``better knowledge than me'' could still make them reopen a decision they had already settled, and FG52 describes this happening repeatedly, as something AI ``wouldn't listen'' to. We take this to suggest that even where judgment survived, participants had to spend something to defend it against a confident AI.

For these participants, relying on AI did not just leave them doubting what they could do themselves; it could also be frustrating: \iquote{P29 -- Survey}{It [AI] was useful but also frustrating as it would often spew out code that did not work correctly.} When we consider these elements together, the frustration, the shame expressed by P19, and the sense of working with a tool that felt like it was gaslighting them, we can see why some participants described feeling disconnected from the project:

\begin{pquotes}

\utt{P5}{But prompting for four months is, like, pointless.}
\utt{P16}{That's a really good point. I feel like I just didn't care. It just was, like, very emotionally disconnected from the work. Like, I was, like, had no pride in it. I was just, like, just...}
\utt{P22}{Yeah, I don't really take much ownership over it either. Like, it doesn't feel like my work.}
\utt{P16}{True.}

\end{pquotes}

\draft{5}{This disconnect recurred among participants who described relying heavily on AI}, though they each put it differently. For some, the work itself came to feel hollow: \iquote{P16 -- Focus Group}{SENG-wise, like, making this, I think I agree with the sentiment that it just felt kind of empty}. However these participants described it, whether as the project being ``pointless'', as feeling empty, as taking no pride in the work, or as not taking ownership of it, the common thread we see is a loss of motivation for the project. We read these accounts as describing participants who no longer saw value in the work they were doing, or a reason to apply themselves to it.

We should be careful not to attribute all of this to AI. Some participants commented that the project itself felt purposeless, and a project a student sees no point in would dampen motivation regardless of how its code got written. Still, we believe the reliance amplified this disengagement, because of what participants attach their disengagement to when they describe it: not the brief or the topic, but ``prompting for four months'', not taking ownership of work that ``doesn't feel like my work'', and the shame and deference discussed above.

One way this showed itself, as P45 recounts, is that participants appear to have stopped caring about the quality of what they were building, letting suboptimal code stack on suboptimal code:

\begin{pquote}{P45 -- Focus Group}
    I found there was a lot of tech debt that you didn't really know was there until you ran into it.
But AI kind of just kept on adding to it and found its own workaround instead of actually fixing the tech debt.
So then it never got done the right way. 
It was just a fix on top of a fix on top of a fix on top of a fix.
\end{pquote}

\noindent This participant does appear to be aware of quality issues in the code, though they also describe technical debt they did not know was there until they ran into it. That AI kept adding to that debt without repairing it, and that the participant let it, reads as a sign that this participant had stopped caring about the project they were working on. This not caring may also connect to the shallow verification discussed by some of the participants in the second insight, as we think the two might reinforce one another. Once a participant no longer cared about the project, why would they carefully verify the AI-generated code or information being added to it? 

Another element we consider part of this insight is the feeling some of the participants had that they were not learning anything. \iquote{P11 -- Survey}{It[AI] is useful for implementing this project, but didn’t help learning.}  We think this feeling, that AI did not help them learn, may have further contributed to some of them disengaging from the project. It matters especially because learning was part of what drew participants in: they told us they were motivated by the chance to pick up new languages and to work with AI. What the project delivered instead, on their accounts, was dependence on a powerful tool that could confidently feed them incorrect information and leave them stuck.

These various experiences that participants shared, of how their AI reliance left them feeling disconnected from the project, do more than describe an unpleasant experience. We think they also help explain why it became even more difficult for participants to break out of their reliance. A participant who finds the project ``pointless'' and takes no ownership of it has little reason to do the harder work of rebuilding their own understanding; it is easier to keep asking AI to make the code. The way P19 talks about the absence of accomplishment hints at this:

\begin{pquote}{P19 -- Focus Group}
    The most of it [excitement] came from just saying, you know, like, magic, make me this code. There is no, there's no sense of ownership, there's no, oh, I am able to create this thing, it's, there's no real sense of pride there.
\end{pquote}

\noindent Because ``magic, make me this code'' produced the project, these participants were denied the experience of ``I am able to create this thing'', the sense of mastery that might have rebuilt their confidence. This loss of motivation, then, did not only feel bad; \draft{4}{it gave participants little reason to break the cycle}, leaving the self-doubt that first pushed them towards AI in place. Some participants described this momentum directly, as a slope they could not get off: \iquote{P6 -- Focus Group}{yeah I guess slippery slope almost like once we started it's like it's already like it's already some areas chopped like you're just gonna keep going.} For us, this is why it matters to consider how the reliance made participants feel. It was not only uncomfortable; it was self-worsening: the less participants cared about the project, the fewer reasons they had to reclaim it from AI, and the more of it they left to AI to produce.

\draft{4}{Across its three insights, this theme traces how participants spoke about a reliance on AI that deepened until they surrendered their judgment, and about the emotional toll of that surrender, which, in our reading, made the cycle even more difficult to break. The second theme begins with what these participants lacked: the knowledge required to judge what AI was doing.}

\subsection{Theme 2: Applying human judgment and the knowledge judgment rests on}

\noindent \draft{8}{The experiences of participants were diverse, and in addition to talking about surrender, participants also spoke about what it was like for them to apply their judgment. This second theme is centered around these accounts.} We constructed this theme around three things participants described: applying their own judgment; the knowledge and experience it rested on, whether brought into the course or acquired during it; and the strategies that kept that judgment in their hands while they worked with AI. One participant states the connection between knowledge and judgment directly:

\begin{pquote}{P36 -- Focus Group}
    Yeah. I think that just points to the importance of you need to actually understand the underlying technologies, tools that you're dealing with, because it just helps with being able to give the right context and problem achieving, as well as understanding if it's somewhat in the realm of a good answer or not.
\end{pquote}

\noindent \draft{1}{For us, this participant summarizes what we construct in the remainder of this theme: understanding the underlying technologies is what allowed participants to give AI the right context and to judge whether what came back was, in the words of P36, ``somewhat in the realm of a good answer''. As mundane as it sounds, a great deal of implicit prior knowledge is required to form such a judgment. The way this participant talks about a ``good'' answer implies a taste that goes beyond asking whether the response achieves the goal, and it speaks to how they cared about the quality of that response.}

\draft{1}{The first two insights of this theme describe where this understanding came from, pre-existing knowledge or knowledge that participants acquired during the course; the third insight describes the strategies participants used to keep it.}

\subsubsection{\draft{5}{Knowing from prior experience what good code looks like}}
\draft{3}{Some of the participants who spoke about applying judgment over AI, also referred to the experience they already had before starting the project. Their experiences reveal to us what it felt to lean on your own knowledge to judge what AI was doing.}

\begin{pquote}{P5 -- Focus Group}
I think I would like all the code that the AI made I reviewed and like make sure that I understood and if there was a problem I knew how to debug it and like fix any problem that there was yeah because I had a co-op in TypeScript and like basically this whole tech stack so I actually kind of knew what was going on
\end{pquote}

\noindent \draft{7}{P5, whom we quoted in the first theme describing AI as ``the only glue'' that let the class progress, gives a different account of their own work when asked whether they vibe-coded their entire project.} That they treat their familiarity with the tech stack as the reason they did not vibe-code it reveals how important they felt their own knowledge was. Moreover, the way the participant states they were able to debug any problems reveals not just familiarity, but also confidence in their own abilities. 

We noticed similar ideas in how P3 discussed their pre-course familiarity with software diagrams and TypeScript: \iquote{P3 -- Focus Group}{even in like previous classes before taking this class I've seen stuff like that [software diagram and typescript] so it was a lot easier to not have to rely on it[AI]} P3 constructs it quite clearly: they did not have to rely on AI because of their pre-existing knowledge, both in software design and in TypeScript, the language required for the project. 

\begin{pquote}{P43 -- Survey}
    I don't know if I'd call it a challenge but part of the procedure for having GenAI create good code requires several iterations and revisions. Everything I had it do for me were tasks I knew how to do myself, so it just accelerated my working speed. When it makes mistakes, I catch them and tell it to fix them.
\end{pquote}

\noindent While talking about using AI for the project, P43 also talks about ``good code'', which suggests that participants like P43 had some ideal, or reference point, for what good code looks like. Developing this kind of taste reveals how this participant used prior knowledge to form judgments about the code AI was producing. This participant only used AI to write code they could judge themselves. 

\begin{pquote}{P44 -- Focus Group}
    I kept prompting it[AI] to do something again and again. I would usually look through what they had given me and see if I can find any problems like what are the specific errors that I'm encountering or what do I not like what about the first way that they did it and then at least for copilot there's a way to restart the context like go back on the context and then I would probably like rewrite the prompt to include what I didn't like or liked about the first one or what I thought I had missed context wise.
\end{pquote}

\noindent While working on the project, we believe this participant had a very specific idea about how the issue should be solved. P44's idea of what should be done went beyond functional correctness: alongside checking for errors, they asked themselves ``what do I not like [...] about the first way that they[AI] did it'', which implies some taste, or subjective ideal, they wanted the code to meet. To make this kind of judgment about the code AI writes, we believe participants must have drawn on their prior knowledge and skills; it is only through prior experience as a developer, we think, that one develops the taste required to be critical of the code AI generates. We found the same in how P37 talks about working with AI: \iquote{P37 -- Focus Group}{Sometimes without doing that [specifically adding context], it[AI] would start trying to make changes in different places that you didn't really want it to.} Both quotes show how involved these participants were with what AI was doing, so involved that they were judging how AI was solving issues. In addition to requiring prior knowledge, these quotes also suggest care about the quality of the project that contrasts with the experiences described in the first theme.

The prior experience commented on by some of these participants might imply there was a clear dividing line between the participants, with the group without prior experience surrendering and the group with experience preserving their judgment. However, we do not believe there is such a clear separation between these groups; there are also accounts of participants who initially surrendered judgment but, throughout the course, restored it. The second insight describes how the latter group did so.

\subsubsection{\draft{5}{Recognizing the gap in knowledge and repairing it}}
\draft{3}{Unlike the participants quoted in the first insight, some participants spoke about the steps they took during the project to acquire the knowledge needed to judge what AI was doing.}

Building that knowledge required space the project did not initially afford participants, and part of that space came from changes to the course. Before the first-phase deadline, students expressed concerns to the course team, explaining they felt they could not meet the deadlines. This prompted the course team to adjust both deadlines, offering a grace period for the first and lessening the requirements for the second. Then, at the halfway point of the course project, immediately after the Phase~I deadline, a lecture and a lab were given to students. The material covered concepts like context engineering and how to apply custom instructions, and it was based on a combination of gray literature articles collected by the course team. \draft{1}{Both were organized around the three-phase workflow described in \Cref{sec:course-context}.} 

Participants commented on these changes in the focus groups \draft{5}{and in the surveys}, which is why we believe they were so important: \iquote{P10 -- Survey}{I relied on it[AI] a bit less because of the reduce[d] workload so I was able to do a bit of research about things rather than just asking it[AI] to do things } This quote shows how important it was for participants to feel like they had the space to research ``things''. In this case, a relatively small change in the conditions was perceived by this participant to afford a different way of working with AI. What P10 says they did is relevant: they were able to research and restore the understanding they had lost during the first phase, and it was through building that understanding that they developed the knowledge needed to apply their own judgment.

Participants also describe becoming aware that the way they had been working with AI was not right, and explain how the course material showed them they could work with AI differently.

\begin{pquotes}
    \utt{P32}{I'd say how to use AI properly is definitely emergent through the course.}
    \utt{FG49}{Yeah}
    \utt{P32}{The professor talking about how to maybe scaffold a way to use it properly actually got me thinking about how to use it properly}
\end{pquotes}

\noindent P32's use of the word ``properly'' implies that, at some point during the first phase, working with AI felt improper, suggesting a moment of self-awareness. While this quote does not tell us whether the lecture triggered this realization or whether it had happened before the lecture, it does suggest that becoming aware of their AI usage was an important element in moving from surrender to applying judgment, as participants who became aware then began to improve their understanding and knowledge. 

When it comes to realizing that they had to improve their knowledge, one team's experience stood out in particular. Just before the first deadline, there was a catastrophic failure that appeared to trigger a moment of realization shared by the entire team.

\begin{pquote}{P41 -- Survey}%This is also a team-six participants
    Before, we just vibe coded to its most extreme level. Didn’t care about outputs, just took it and ran. Once that failed spectacularly, we switched to using it much more sparsely and cautiously.
\end{pquote}

\noindent Based on what the members of this team told us, it is apparent they were unable to continue working on their project after the first deadline. It shows how the structure of the course was instrumental for some students; without the initial tight deadlines, the participants from this team might not have ended up surrendering. While P41 did not refer to the change in course deadline, we think the change must have played a role in helping these participants recover their judgment. For P41 to refer to this switch as ``using it[AI] much more sparsely and cautiously'' again implies that this participant was actively aware of how the initial reliance had been problematic. The failure is hinted at as a catalyst, showing how this participant constructed boundaries around their AI usage in response, leading to an internalized notion of what it means to use AI sparingly and cautiously. 

Recognizing the problem was not by itself enough to work differently. What this team did next is the clearest example in our data of participants deliberately acquiring the knowledge required to apply judgment, which this insight rests on:

\begin{pquotes}

\utt{P24}{uh for team [redacted] for implementation II we all spent three hours out of our day and watched a video on nestjs and came to the conclusion that AI is not good for helping you with coding}
\utt{}{...}
\utt{P24}{And then we came to the conclusion
that the AI cannot do NestJS complete.}
\utt{P41}{Yeah.
It needs a bit more structure.}
\end{pquotes}

\noindent In the middle of a project with demanding deadlines, all team members committed to watching the same series of YouTube tutorial videos on the framework used in their project, spending ``three hours out of our day'' on it. What makes this notable is what the knowledge was for: it was only after learning the framework themselves that the team ``came to the conclusion that the AI cannot do NestJS complete''. Until then, we think they were unable to judge the code AI produced for this framework. \draft{1}{That a relatively short series of videos was enough also reveals the importance of the prior software engineering knowledge these participants already held.}

\draft{1}{Two further things stand out to us. For this team, acquiring the knowledge was not an individual action: making it a group activity shows the whole team thought it important that everyone could apply judgment to what AI was doing. And the way participants speak about the realization implies that they recognized not only that AI has limitations, but, as P41 alludes to, that it is their own responsibility to give the models the structure they need.}

The same acquisition of knowledge and experience happened more quietly for other participants. Before starting the project, many participants did not feel very familiar with web development; throughout the project, however, this started to change: \iquote{P8 -- Survey}{I tried to use it[AI] less in phase 2 as I felt slightly more confident with the TypeScript basics} This quote shows how becoming more familiar with the web technologies required for the project was instrumental for this participant in reducing their AI reliance. However, it also suggests that, during Phase I, this participant started to build some of the experience required to judge AI output. 

In addition to becoming more experienced with web development, some participants also commented on the familiarity with AI they developed as the course progressed. Even though participants mentioned they had worked with AI tools before the course, for most of them using AI this intensively for software engineering was a first.

\begin{pquote}{P6 -- Survey}
    I got more proficient using [AI]. I was able to manage the scope of change much more fluently, and move slow with AI, but with higher quality codes. I found myself using less [AI], and wanting to write more of the code by my own hands
\end{pquote}

\noindent P6 highlights how the feeling of becoming more proficient with AI gave the participant the space to ``move slow''. Like the quote from P10, we think this shows how the participant constructed working with AI: there is a trade-off where uncontrolled AI usage was quicker, helping participants achieve goals for the project (a functional system), but at a cost to their own ability to judge, and at a cost to quality. As we have seen in the first theme, some participants hinted at being aware of quality problems in their code, so it makes sense that participants felt moving more slowly allowed them to improve quality. What this felt like for P6 is worth noting: as they became more proficient, they describe ``wanting to write more of the code by my own hands'', which reads as the opposite of the emotional distance from their own project that participants described in the first theme. It suggests engagement and enjoyment in doing the work themselves, instead of surrendering \draft{8}{their judgment to AI}. 

\begin{pquote}{P41 -- Focus Group}
    So I said, AI was like milk and a cup of tea. A little bit would enhance the drink, but too much could ruin it and require you to remake it.
\end{pquote}

\noindent In their closing metaphor, P41 describes what we consider to be a core element of this realization participants had. Used purposefully and carefully, AI enhances their work, but overuse has serious, even irreversible, consequences. Knowing there is a ``too much'' implies that participants discovered a boundary around their AI usage, and it shows how participants of this team appeared to care about the quality of their code, actively sought information to restore their judgment, and were able to dose the AI.

\subsubsection{\draft{5}{Dosing the milk --- strategies that kept judgment in participants' hands}}
\draft{1}{When participants applied their own judgment, they did so through strategies for working with AI that they discovered during the project, or in the words of P41, they dosed the milk. These strategies are how participants put the judgment described in the previous insights into practice. What they have in common is what they protect: the chance to notice when AI was going the wrong way, a unit of work small enough to check, the decision about what to build, and the participant's own understanding of the code. We believe that, beyond their practical value, discovering these strategies also restored some of the confidence in their own work that participants described losing in the first theme.}

\draft{4}{We consider these strategies noteworthy because they are what worked for this cohort. Unlike senior developers, these participants could not draw on years of knowledge and experience, so the ways senior developers offload large parts of an implementation to AI are unlikely to have worked for them. Instead, they discovered strategies that suited their own knowledge and experience, and their mastery of working with AI.}

For some participants, how they had learned to work with AI was difficult to put into words:

\begin{pquotes}
    \utt{FG48}{Like, whenever it would start to go off track, I'd just kill that chat and start a new one.}
    \utt{Comment}{P45 and P42 nod.}
    \utt{Moderator}{And when you say off track, you mean?}
    \utt{FG48}{You just kind of can feel that it's going the wrong way.
I don't know how to sort of explain that, but do you?}
\utt{P45}{For me, it just keeps on giving the same answer without actually proposing a fix.}
\utt{FG47}{It's, like, poisoning its own context.}
\utt{FG46}{Or, like, when you're trying new solutions and it just, like, doesn't work.
Yeah, the same reason, maybe.}
\end{pquotes}

\noindent To us, this focus group interaction illustrates how important a sense of intuition was for the participants. Several of the participants appear to find it hard to name the signals they were paying attention to, instead only commenting that they could ``feel'' AI get off track. When prompted to explain this, participants not only refer to what we would consider intuition; they also appear to have discovered terminology, such as ``poisoning its own context'' (FG47), to describe the AI behavior they had become familiar with. Moreover, for P45, an important signal of AI going off track is it giving the same answer without proposing a fix. This is even more notable because, in the first theme, a participant encountered the same problem but did not interpret it as a signal to adjust their use.

\begin{pquote}{P37 -- Survey}
    There was still a heavy reliance on the use of gen AI due to the necessity to produce so much code in such a short amount of time, however, I found it more effective to maintain smaller chats and having only the necessary context for the immediate problem I tried to solve. This helped generate better code results. As well, using instructions helped better control the AI in providing some of this necessary context.
\end{pquote}

\noindent This quote is a response by P37 to a prompt asking whether anything changed in their AI usage from Phase~I of the project to Phase~II. It shows two things. First, even though the deadlines for Phase~II were changed, the participant still experienced pressure, as evidenced by how they speak about having to produce so much code in a short amount of time. Second, the quote shows that this participant appeared to feel more capable after discovering strategies for working with AI. Both of these strategies, maintaining smaller chats and using custom instructions, are very specific approaches to context engineering\footnote{\url{https://www.promptingguide.ai/guides/context-engineering-guide}}. We think this quote shows how the participant experienced a sense of control while working with AI.

\draft{1}{We want to name P37 as a case that does not fit this theme cleanly. The strategies restored a sense of capability, but on this participant's own account neither the deadline pressure nor the ``heavy reliance'' itself went away. Judgment, here, is something the participant regained \emph{within} a continued reliance on AI, rather than something that brought that reliance to an end. We return to what this implies in \Cref{sec:discussion}.}

\begin{pquotes}
    \utt{P4}{i normally uh would use the co-pilot like built into vs code and i would make sure that it looks through documentation like i pointed towards the documentation requirements i need for a feature um i'll tell it to do an incremental steps so i can test it in between each like small portion of this uh requirement it needs to do }
\utt{P9}{ i was gonna say like as well like on top of that i feel like i did a similar thing like asking to do incremental and then also kind of like explaining or like making a plan first so it's like asking it to make a plan and also like explaining the reasoning and then i could say like okay actually like let's maybe do this instead like i'm kind of editing the plan if that makes sense}
\end{pquotes}

\noindent Similarly, the two participants in this interaction refer to several specific strategies when asked how their AI usage changed. P4 describes instructing the AI models to consult the project documentation, another example of context engineering, but also having AI work in incremental steps that can be tested one at a time. P9 elaborates, describing how they had AI generate a plan first, which they could then manually tweak. The two strategies protect different things: the incremental steps keep the work small enough for the participant to check, whereas editing the plan keeps the decision about what to build with the participant. We read these strategies as showing that, throughout the course, participants not only became more familiar with web development; they also learned to use AI in a way that let them exercise their judgment: small, testable steps and editable plans gave them exactly the room to notice and correct what AI got wrong.

Not all strategies were aimed at controlling what AI produced. In the following exchange, P12 and P19 describe strategies aimed at understanding the code produced with AI:

\begin{pquotes}
    \utt{P12}{and I wanted to know what each line of code was doing for that specific feature.}
    \utt{P19}{what he [Points to P12] said. I was just asking [AI] to write more comments in the code and explain every step rather than just ``write me this thing''. }
\end{pquotes}

\noindent Both P12 and P19 refer to using AI in ways that helped them preserve their understanding of the code. While they do not comment on it explicitly, we believe this preserved understanding is exactly what participants needed to keep exercising the judgment described in the previous insights. Where the first two insights describe knowledge participants brought or acquired, this strategy is about not losing it again.

\draft{4}{Both themes come from what participants told us \draft{5}{in the focus groups and in their open survey responses}, and together they describe two ways of working with AI: one in which participants judged what AI produced, and one in which they could not, or did not. \draft{5}{Participants' answers to the closed survey questions, given before and during the project,} give little sign of either experience.}

\subsection{Self-efficacy, and AI perceptions of the participants}\label{sec:results:descriptive}

\begin{figure}
    \centering
    \includegraphics[width=0.95\linewidth]{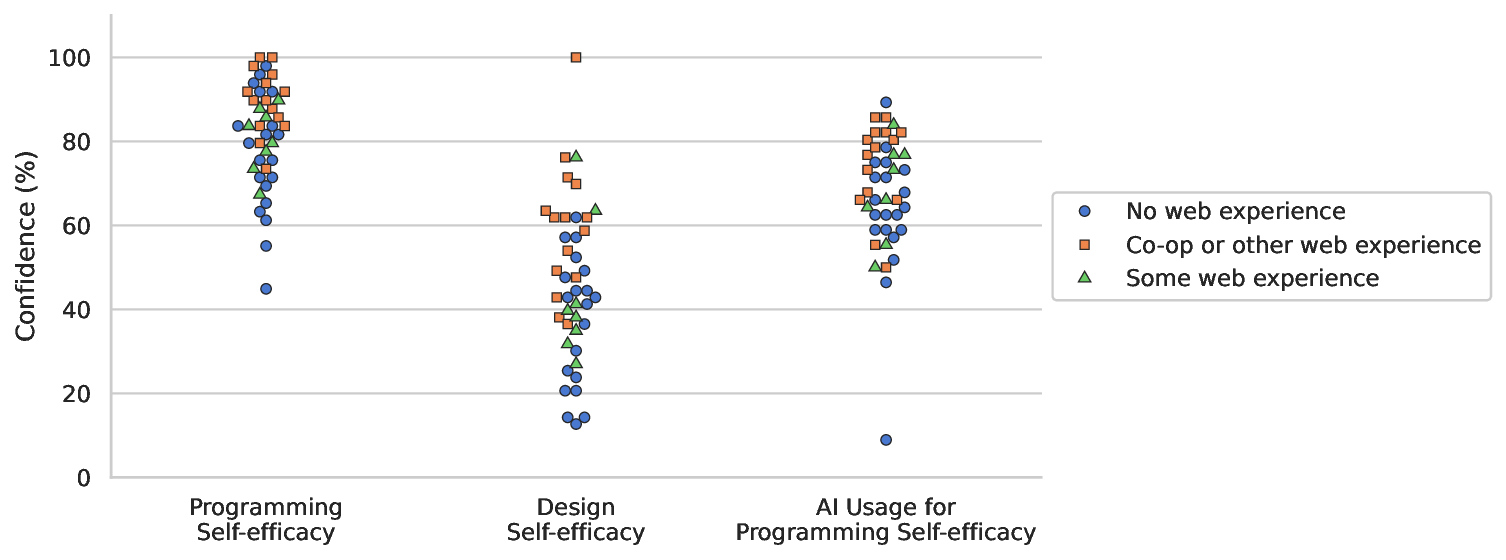}
    \caption{The self-efficacy scores of the participants for the three knowledge areas (Programming, Design, AI usage for programming) divided by the prior experience of participants. }
    \label{fig:self-efficacy-scores}
\end{figure}

\draft{3}{In addition to the focus groups, we also deployed three surveys to participants throughout the project. The surveys contained a set of closed questions covering prior experience, self-efficacy, and the AI perceptions of participants. In this subsection we discuss those results, and how they relate to the themes we constructed \draft{5}{from the focus groups and the open survey responses}.  }

In total, 43 students participated in the self-efficacy survey at the start of the course. Their prior experience varies widely: 20 participants reported no experience with web development, 8 reported some experience, and 15 reported more extensive experience through co-ops or other web development roles. \draft{1}{The survey item offered four response options. We merged two of these, having worked on a web application during a co-op or other term project, and free-text ``other'' responses describing comparable web development work, into the single ``more extensive experience'' level reported here.} 

In terms of self-efficacy, or the confidence participants reported in their own skills, \Cref{fig:self-efficacy-scores} shows the scores reported by each participant for the three categories (Programming, Design, and Programming with AI). A score of 100 indicates that, before the start of the course, a participant was very confident they could solve various tasks in each category, without AI assistance.

Taken together, these prior experiences and self-efficacy scores show that almost all participants reported high confidence in their independent programming skills, even those with limited prior web development experience. They also show that most participants felt confident in their software design skills, \draft{1}{likely from prior courses on design patterns}. Based on these prior experiences and self-efficacy scores, it appears that participants had the skills and confidence the teaching team expected. We do note, however, that this confidence concerned participants' general programming skills, not their confidence in the specific web development stack the project required, which nearly half of them reported no experience with.

After each development phase, participants completed a survey on their AI usage and perceptions (37 and 38 responses, respectively). Except for one student, who reported after both Phase~I and Phase~II that they had not used any generative AI,\footnote{This participant gave two responses when asked why they did not use AI: ``\emph{I’m in this program because I love coding and software and understanding how it all works. Generative AI is an easy way to get some code to exist (in whatever quality the AI spits out) but it does nothing to help me actually understand and learn. I think it’s unfair to essentially require students to use a gen’ AI to complete a project on time. If some students want to use AI to make it easier, let them, but don’t expect everyone to use it. I’m here to learn the material myself. Once I understand what’s going on then I can benefit from gen’ AI. Thank you :)}'' \& ``\emph{I want to understand my teams code. I do not understand code without working through it myself. I want the documentation I write for my team to reflect my understanding, even if I could fake that understanding with genAI. GenAI makes me work faster at the expense of being able to understand what I am doing.}''} all students indicated that they used generative AI, similar to other courses where students are allowed to use AI~\cite{Salomon:2025, Wang:2026}. 

\begin{figure}
    \centering
    \includegraphics[width=0.95\linewidth]{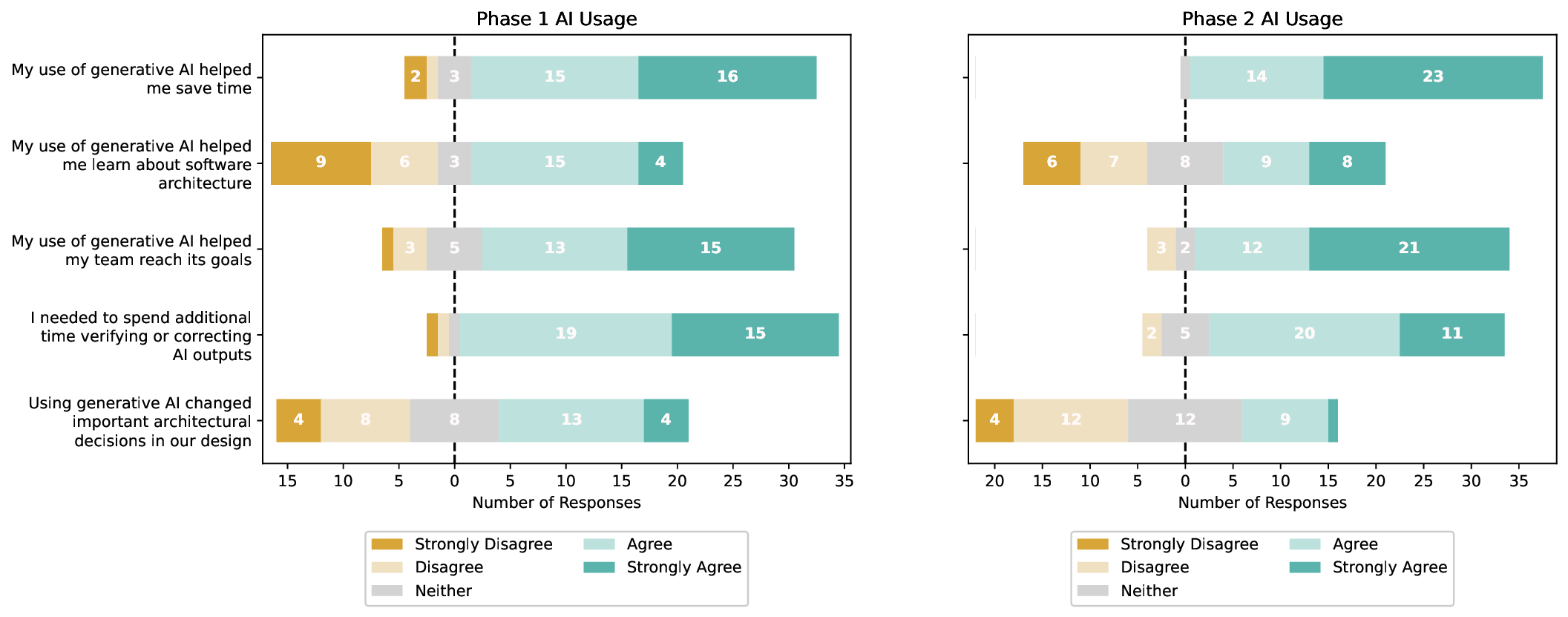}
    \caption{Participant perceptions of AI after Phase I and Phase II. }
    \label{fig:ai_usage}
\end{figure}

\Cref{fig:ai_usage} shows participants' perceptions after each phase. Almost all participants reported that generative AI helped them and their teams reach their goals or save time. These perceptions suggest that, after each deadline, participants believed AI was effective and that it helped them complete the course project. This near-universal adoption and perceived usefulness is consistent with what large-scale surveys in comparable project-based software engineering courses report~\cite{Salomon:2025}. At the same time, it is notable that participants were less positive about how generative AI affected their learning, with 15 and 14 participants, respectively, disagreeing with the statement that AI helped them learn software architecture.

\draft{4}{What stands out to us is that these \draft{5}{closed} survey responses give little sign of the surrender, or of the emotional toll that came with it; at most, the doubts about learning echo P11's sense that AI ``didn't help learning''. This suggests that surveys like ours are not suited to detect when students surrender their judgment, or what that surrender costs them. This is why we think an experiential account of AI adoption matters: it shows what working with AI is like for software engineering students, an experience that perception surveys have so far missed.}

%% file: sections/6-discussion.tex
\section{Discussion}
\label{sec:discussion}

In the focus groups, participants shared what it was like to work with AI. \draft{8}{Some accounts revealed a disheartening experience: a deepening cycle in which judgment was surrendered to AI (\emph{Surrendering Judgment to AI}). While speaking about surrender, participants also spoke of shame, frustration, and an emotional disconnect from their projects, a toll that, in our interpretation, made the cycle even harder to break. At the same time, participants also spoke about what it was like to apply their judgment to AI,} how that judgment rested on their knowledge and experience, and how they cared about what they built (\emph{Applying Human Judgment}). We believe our findings help uncover what the next generation of novice developers needs in order to adopt AI sustainably, and what must be done to preserve the motivation and well-being of junior developers. 

\draft{5}{In the remainder of this section, we use Self-Determination Theory as a lens to explain why the reliance on AI left participants feeling disconnected and demotivated. We then discuss what our findings imply for software engineering education and for junior developers in industry, and argue that both settings need diagnostics that detect when students and developers surrender their judgment to AI.}

\subsection{\draft{5}{Why surrender felt empty: interpreting the findings through the lens of Self-Determination Theory}}\draft{5}{\label{sec:discussion:sdt}}

We constructed the themes inductively, based on the excerpts and codes. However, as encouraged by \citet{Braun:2022}, we were open to theories that could help interpret and contextualize our themes. Specifically, we believe Self-Determination Theory (SDT)~\cite{Ryan:2000} helps explain how participants came to describe their work as empty and hollow. \draft{4}{We also considered Flow~\cite{Csikszentmihalyi:1990} and Cognitive Load Theory~\cite{Sweller:1994}, but found SDT the most useful for explaining the emotional disconnect that participants repeatedly described.}

SDT is an established theory centered on human motivation and the conditions that enable humans to thrive. \draft{2}{According to SDT, people have an inherent tendency to grow, but this tendency needs supportive conditions, and can be readily disrupted without them~\cite{Ryan:2000}. SDT further posits that motivation is not one quantity but a continuum of regulatory styles, ordered by how far the reason for acting lies within the person themselves. At one end sits amotivation, the state of lacking the intention to act, where people either do not act at all or go through the motions. At the other end sits intrinsic motivation, doing an activity for its inherent satisfaction. Extrinsic motivation covers the range between the two: acting not for the activity itself, but for an external reward. Its regulatory styles differ in how autonomous they are, from external regulation, where someone acts to obtain a reward or to satisfy a demand, through introjected regulation, where someone acts to avoid guilt or to protect their self-esteem, to identified and integrated regulation, where someone performs a task not because it is fun, but because they hold it to be personally important. SDT has been applied to a variety of domains, including education~\cite{Ryan:2000}, and it has been used to understand how developers experience their work~\cite{Xin:2024}.}

SDT states that three basic needs must be satisfied for someone to stay motivated and to thrive: \textbf{Autonomy}, \textbf{Competence}, and \textbf{Relatedness}~\cite{Ryan:2000}. Autonomy is the need to experience one's actions as volitional and as one's own. Competence is the need to feel effective and capable of meeting the demands of a situation. \draft{4}{Relatedness is the need to belong: to feel connected to others and cared for by them.} Based on the experiences participants shared in Theme 1, we believe both their autonomy and their competence were thwarted.

\draft{2}{According to SDT, thwarting these needs is likely to undermine an individual's motivation. \citet{Ryan:2000} state this as a property of the environment rather than of the person. SDT places an individual at amotivation, the lowest level of motivation, when they do not feel competent to do an activity, do not value it, or do not expect it to produce the outcome they want. For instance, an individual might take a course because they want to learn a skill. However, if the course is too difficult, and the individual does not feel competent, they might come to work at just passing the deadlines of the course rather than at learning the skill (external regulation), and, if that continues, stop seeing any point in the work for the course at all (amotivation).}

\draft{2}{\draft{5}{Determining the precise extent to which participants in this study felt motivated is not possible. The focus groups asked participants about their motivation (\Cref{tab:focus-group-script}), but we did not use any validated instrument to measure it.} However, some participants' quotes about the start of the course suggest that participants began with a relatively high level of motivation. For instance, P18 said they were motivated to learn web technologies. We do not read this as intrinsic motivation. P18 wanted to learn TypeScript because it is widely used in industry, \draft{8}{which does not suggest the only motivation for this participant was learning typescript for the sake of TypeScript itself.} SDT calls this identified regulation, where someone does the work because they believe it is important~\cite{Ryan:2000}. We think this fits a course project, as \citet{Ryan:2000} note that people are only intrinsically motivated for activities they find interesting in themselves. As the course continued, participants describe conditions that, in SDT terms, would thwart their needs, and we hypothesize this undermined their motivation.}

\draft{2}{For autonomy, we think the structure of the project deadlines left some participants feeling like they had little control over their own actions. \citet{Ryan:2000} already present deadlines and imposed goals as factors that undermine motivation, because they move the reason for acting outside the person. Several participants referred to the pressure they felt from the deadlines directly (P9, P37). \draft{8}{Deadlines alone are not sufficient to thwart this need, but feeling overwhelmed, as some participants described being, would.} On top of autonomy, we also believe these participants did not feel competent. P22 called AI a ``massive crutch'', and said their group could not have progressed without it. They also told us that ``None of us felt like we knew more than the AI did'', which suggests that some participants did not feel competent enough to complete the work without AI. }

\draft{2}{In light of SDT, we think how participants experienced their sense of autonomy and competence might explain why some participants came to describe their work as hollow, empty, or felt emotionally disconnected. \draft{8}{We believe these accounts hint at a move towards external regulation}, where the only reason left to act is to pass the project rather than to learn or to grow, and for some participants as a move towards amotivation, where they no longer saw a reason to act at all. We should be careful here, because participants kept working and all teams delivered a system. We therefore do not claim that participants fell into complete amotivation, but we do think our findings suggest that participants that surrendered to AI started to experience reduced motivation.}

\draft{4}{Our instruments did not cover relatedness, but prior work suggests that AI use can affect it too. Students who use AI report feeling less connected to their peers~\cite{Choudhuri:2026,Salomon:2025} and to their instructors~\cite{Salomon:2025}. Our data hints at the same substitution. FG47 preferred going to AI over ``asking a teammate or a TA or prof'' because of the time constraints (\Cref{sec:findings}). When students turn to AI instead of to their peers and instructors, we think this may also thwart their need for relatedness and further undermine their motivation.}

Because we adopted SDT as a lens after we constructed the themes, \citet{Braun:2022} also suggest discussing how the theoretical lens might be limiting. There are two aspects we discuss. The first is relatedness, the sense of belonging, which is one of the three needs identified in SDT, but which we did not extensively cover in the focus groups. \draft{2}{Of the three needs, autonomy and competence are the ones SDT ties most directly to motivation. \citet{Ryan:2000} report strong links between intrinsic motivation and the satisfaction of these two needs, and describe relatedness as bearing on motivation in a more distal way. They also note that people happily do intrinsically motivated work on their own. We therefore think our account remains consistent with SDT, as the emptiness described by participants might be explained by their lack of autonomy or competence. The second is the role of the team. Several accounts discuss the team: P5 describes AI as the only ``glue'' that let their team move forward, and the team of P24 and P41 reclaimed their judgment collectively, by all spending three hours watching tutorials on the framework they were using (\Cref{sec:findings}). Our findings do not extend to, or discuss, the team dynamics.}

\draft{3}{Regardless of these limitations, we believe SDT is a good lens for understanding how the course and the deepening cycle of AI reliance affected participants' motivation. It helps explain why participants came to describe their work as hollow, and it grounds the implications we draw from this study.}

\subsection{Implications for software engineering education}\draft{5}{\label{sec:discussion:education}}

\draft{4}{The first set of implications that follows from this work concerns education. \draft{5}{Existing literature links AI overreliance to cognitive debt, which \citet{Kosmyna:2025} measure as reduced neural engagement during AI-assisted tasks and \citet{Storey:2026} describes as a team's eroding understanding of its system. Others speculate that overreliance might lead to skill atrophy~\cite{Choudhuri:2026}. To prevent these outcomes,} prior work proposes AI literacy: making students aware of the limitations of AI and teaching them to work with it~\cite{Choudhuri:2025:Trenches,Choudhuri:2026,Salomon:2025,Guner:2025,Sengul:2024,Anderson:2026:Stuck}. Our findings suggest that AI literacy is only part of the picture: participants whose accounts we report in Theme 1 knew AI could give them incorrect information and accepted that risk anyway (\Cref{sec:findings}), so awareness alone did not prevent the reliance that left them emotionally disconnected. Likewise, agent skills or custom prompts that claim to adapt AI models to support learning do not address any of the conditions that we believe led students to surrender their judgment. Participants described working with a technology they felt knew more than they did, and that at times felt like it was gaslighting them; teaching them how to work with it, we think, only partly addresses that experience. \draft{7}{Two further conditions matter: the knowledge students have before they use AI, and the environment in which they learn.}}

\draft{4}{First, students need relevant knowledge before they work with AI, because without it they have nothing to ground their judgment on. \draft{7}{The strategies participants described in Theme 2, such as starting fresh chats, writing custom instructions, editing the plans AI generated, and working in small, testable steps, are worth teaching, but each rested on knowledge of the stack that participants brought or acquired.} Our participants were senior students, yet many worked with a stack they did not know (20 of 43 reported no web development experience), and the ones who reclaimed their judgment did so by first building the knowledge they lacked (\Cref{sec:findings}). We therefore think it may be better to withhold AI from students, even in senior software engineering courses, at least until they have built the knowledge a project requires. Restricting AI is not a new idea: some instructors have banned generative AI in their classes, and while students disagree with outright bans, a majority support some restrictions on its use~\cite{Rahe:2025}. \citet{Choudhuri:2026} propose a milder form, in which students alternate between AI-assisted work and supervised work without AI, so that the cost of delegating their thinking becomes visible.}

\draft{4}{Second, the environment in which students learn matters. Participants described deadlines that left them little control over how they worked, and some told us the relaxed deadlines gave them room to research rather than delegate (\Cref{sec:findings}). We think students whose sense of autonomy is undermined in this way are more likely to surrender their judgment to AI, and because AI can complete much of their coursework, that surrender may go unnoticed. \citet{Abbas:2024} argue that an environment in which students derive satisfaction from mastering concepts independently mitigates overreliance.}

\draft{4}{Read through SDT, restricting AI and shaping the learning environment protect the same thing: the competence and autonomy students need before AI can support their work rather than replace their judgment. It is therefore imperative that software engineering education is designed so that students never reach the point where they surrender their judgment, because the slide towards amotivation that follows frustrates the development of the skills they need in their degree and future careers. \citet{Ryan:2000} report that more autonomous forms of motivation are associated with higher-quality learning, better performance, and more engagement. That is what students who surrender stand to lose.}

\subsection{The vulnerable position of junior developers}\label{sec:discussion:juniors}

This research project studied the experiences of students in a software engineering degree program. However, based on reports of AI adoption in industry, we think junior developers might be at risk of the amotivation the students described. Three of the conditions we observed in this study also hold for many junior developers entering industry: an unfamiliar stack, delivery pressure outside of their control, and an expectation to use AI to enable productivity gains. 

The first two conditions are long-standing. Junior developers are often asked to work in languages or frameworks they do not yet master, and the demanding deadlines in industry have been extensively documented~\cite{Xin:2024}. The third condition is more recent. \citet{Miller:2025} describe a so-called ``productivity paradox'', where developers who received access to AI tools were expected to be more productive. Not all of these developers were immediately able to use AI effectively, and they described feeling more pressure to be productive. Read through the lens of SDT, these accounts suggest to us that the expectations around AI might undermine the autonomy of these developers in much the same way. The three conditions also bear on competence: junior developers starting their first job in industry have long reported feeling insecure when working with unfamiliar technologies~\cite{Begel:2008}. \draft{2}{Interpreted through SDT, this insecurity might be a sign that their need for competence is thwarted.}

\begin{table}[tp]
    \centering
    \begin{minipage}{\linewidth}
        \setlength{\columnsep}{8pt}
        \raggedcolumns
        \begin{multicols}{5}
            \input{sections/metaphors}
        \end{multicols}
    \end{minipage}
    \caption{\draft{6}{The closing metaphors of the 31 participants who shared one (\Cref{tab:focus-group-script}), ordered by participant identifier down each column. Darker notes mark the metaphors we discuss in the text. Wording in quotation marks is verbatim; the remaining entries are condensed from the transcripts.}}
    \label{tab:metaphors}
\end{table}

How these conditions might affect junior developers is visible in how the participants came to describe AI. \draft{6}{At the end of each focus group, we asked participants for one or two words, or a metaphor, for the role AI had played in their project (\Cref{tab:focus-group-script}). \Cref{tab:metaphors} lists all metaphors. Many are positive, and those that cast AI as a solution or a time saver echo the survey responses (\Cref{fig:ai_usage}). Several, however, cast AI as knowing more than the participant: a mentor, a subject expert, a ``god'' (P12, P16). Others turn on its confidence: ``an overconfident friend'' (P8), a friend who has ``been manipulating you'' (FG53), an ``intelligent gaslighter'' (FG51). Three participants cast AI as a junior developer, but P45's junior developer knows everything.}

\draft{6}{Together, these} metaphors reveal how the participants, who were still novices, came to view AI throughout the project. \draft{2}{They show how participants made sense of the characteristics of AI, such as it being confidently incorrect~\cite{Zhou:2024}, and how some came to see AI as an all-powerful knowledge base. To us, these metaphors further suggest that the participants who surrendered did not feel effective while working with AI. On top of the amotivation we described, this suggests that after they \draft{6}{have surrendered their judgment to AI, juniors are left working with a tool they do not feel able to judge}.} We believe this is not a desirable situation, and therefore we pose the following question, which we believe is most important for the future of junior developers: 

\vspace{.5em}

\framingquestion{\draft{1}{How do we help developers and organizations adopt AI sustainably, to preserve junior developers' sense of craft and ownership?}}

\vspace{.5em}

\noindent Existing reports about AI adoption in industry already reveal how developers might be at risk. In a longitudinal study, \citet{Vella:2026:Longitudinal} find that over time, 25\% of developers reported that AI was reducing their developer experience. In August 2026, a video in which Brett Codes discusses why they stopped using AI for coding went viral on YouTube.\footnote{\url{https://www.youtube.com/watch?v=2ZU3j4GQ4K8}} Responding to this video, Gergely Orosz, author of \emph{The Pragmatic Engineer}, reported that they had also heard signals of developers burning out because of AI.\footnote{\url{https://x.com/GergelyOrosz/status/2088567020623151475?s=20}}

One of the reasons Brett Codes gives for stopping is very similar to some of the accounts given by the participants in our study: ``Using these AI agents for coding creates this apathy towards the work and the software. At least that's what I felt. Where I just care significantly less about things the AI generated.'' Unlike our participants, Brett Codes has 19 years of programming experience. At the same time, we believe they describe an experience very similar to what participants in this study went through. \draft{1}{In their account, Brett Codes also describes external pressure: they felt pushed to be more productive with AI, and were mandated to use it. } That even an experienced developer reports such an emotional disconnect shows us \draft{2}{that the surrender we observed might not be limited to this study.}

As relevant and important as we consider this question, we think that the tools and methods currently proposed to help junior developers adopt AI sustainably fall short. 
\draft{5}{In \Cref{sec:discussion:education}, we argued that AI literacy, which prior work proposes for students~\cite{Choudhuri:2025:Trenches}, addresses only part of what led participants to surrender. Proposals aimed at industry share this focus on competence: \citet{Kam:2025} state that professional developers need to master more than working with AI, and also need software engineering skills. This recommendation is relevant.} However, our \draft{5}{findings}, interpreted through the lens of SDT, \draft{2}{may} suggest that a focus on competence alone is not sufficient to prevent amotivation. The findings of \citet{Choudhuri:2026} already hint at this, as they conclude that for some students, an active awareness of the limitations of AI does not in itself deter a reliance on AI. 

The participants in our study who reclaimed their judgment did so after the conditions of the project changed. The course team relaxed the deadlines and taught a lab on working with AI, and participants' accounts suggest that both mattered: some told us the course material showed them they could work with AI differently, and P10 told us the reduced workload was what gave them room to research rather than to delegate (\Cref{sec:findings}). P6 describes that room from the other side, as being able to ``move slow with AI, but with higher quality codes''. \draft{2}{In SDT terms, the two changes addressed different needs: the lab supported competence, and the relaxed deadlines supported autonomy. Both might therefore have changed how participants regulated their motivation while they worked with AI.} Our results suggest how important these two different interventions were in helping participants reclaim their judgment. \draft{2}{What we take away from this is that teaching juniors to evaluate AI output might not be enough to help them reclaim their judgment, if the conditions they work under leave them without a sense of autonomy.} Of the three conditions we described in this study, delivery pressure and AI usage mandates are the ones an employer controls most directly.

We hope the accounts we report help organizations and junior developers adopt and work with AI in a way that preserves the craft and motivation of junior developers. As junior developers are expected to acquire new skills through their work, it is important they work in environments that facilitate that. \draft{4}{Whether developers will still write code by hand is an open question in industry,\footnote{\url{https://newsletter.pragmaticengineer.com/p/when-ai-writes-almost-all-code-what}} but we do not think the answer removes the need to learn on the job. Our participants worked with the AI tools available in 2025, and those who had AI generate their code still got stuck; those who got unstuck on their own terms first built the knowledge they lacked. We therefore think junior developers will still need to build knowledge and experience through their work, whatever direction software engineering takes. Like \citet{Miles:2026}, we believe juniors must keep developing their craft, because that craft is what allows them to apply judgment.}

\draft{4}{We worry that juniors who surrender their judgment to AI lose both the ability and the will to develop that craft. Once they surrender, AI makes the decisions for them, they stop evaluating its output, and they start to feel disconnected from the work. Because more controlled forms of motivation are associated with worse learning outcomes~\cite{Kusurkar:2013}, we think these juniors will also acquire new skills less effectively. That is why we think the question we pose matters: answering it can help keep the next generation of developers from becoming dependent on AI, losing their motivation, and never building the knowledge and craft they need to judge AI and to grow into independent, senior developers.}

\subsection{\draft{8}{Developing the ability to detect surrender}}\label{sec:discussion:diagnostics}

Both sets of implications depend on being able to notice when students or junior developers surrender their judgment to AI, and in this study the surveys, and especially the closed questions within those surveys, were not the right tool to detect this. Before the project began, almost all participants reported high confidence in their own programming skills (\Cref{fig:self-efficacy-scores}). After each phase, almost all of them reported that AI helped them and their teams reach their goals or save time (\Cref{fig:ai_usage}). With similar perception-based instruments, \citet{Salomon:2025} report much the same surface picture at a larger scale, where 84\% of students used generative AI and over 90\% of them found it useful. \draft{5}{None of these measures surfaced what participants told us in the focus groups and the open survey responses.}

We therefore argue that educators and organizations need better diagnostics to help identify whether students and developers are in the process of surrendering their judgment. \draft{8}{In particular when it comes to the emotional disconnect that participants spoke about when they spoke about surrender. In fact, the account shared by Brett Codes suggests this may already be happening to developers outside of this course, and the work on cognitive surrender in software engineering education further emphasizes how this might be a widespread problem~\cite{Choudhuri:2026}. Therefore, we believe that the accounts we report contribute a starting point: the words participants used to describe how surrender felt could seed instruments to further understand the concept of surrender, and afford us the ability to detect it early.}

Until such instruments exist, the design of our focus groups offers one way to surface surrender (\Cref{tab:focus-group-script}). Each session anchored the discussion in a concrete architectural decision or feature from the project, invited two teams to build on and challenge each other's accounts, and closed by asking each participant for a metaphor for the role AI had played. Key accounts in this paper came from both the exchanges and the closing exercise: the disconnect P16 and P22 described surfaced as they built on P5's remark that prompting for four months was ``pointless'', and the metaphors of AI as a gaslighter came from the metaphor question (\Cref{sec:findings}, \Cref{tab:metaphors}). Educators and researchers can reuse the elements from our design in their own courses, together with the survey instruments and course materials that are made available in our replication package.

\draft{5}{Diagnostics are also the first step to answering the question we posed in \Cref{sec:discussion:juniors}. In our course, the course team relaxed the deadlines and taught a lecture and a lab on working with AI partway through the project, and participants' accounts suggest these changes helped some of them reclaim their judgment. Diagnostics could show educators and organizations when such changes are needed, and for whom.}

%% file: sections/metaphors.tex
%% Closing metaphors shown in tab:metaphors (sections/6-discussion.tex).
%%
%%   \metaphor{<id>}{<text>}    one sticky note
%%   \metaphor*{<id>}{<text>}   darker note: a metaphor the discussion text cites
%%
%% Wrap verbatim wording in ``...''; leave condensed wording unquoted.
%% Entries quoted elsewhere in the paper use the paper's wording.
%% Notes flow in this order down each column, then into the next column.

\metaphor{P3}{A solution}
\metaphor{P4}{A taxi driver from a small town on the west coast, thrown into downtown New York}
\metaphor{P5}{Trying to cross the ocean without a boat or a plane; AI is what helps you cross, and the project would not have been possible without it}
\metaphor*{P8}{``an overconfident friend [...] who sometimes gives you a wrong answer''}
\metaphor{P9}{A gateway to information; help, for the most part, but sometimes it trips up and is wrong}
\metaphor{P11}{It did everything; a 100x software engineer}
\metaphor*{P12}{A ``god''}
\metaphor{P14}{Incredibly helpful}
\metaphor*{P16}{A ``god'', an ``all-powerful subject matter expert''; dulling, and overpowered}
\metaphor{P18}{A junior developer: as if I'm a senior developer overlooking a junior's changes to see whether they are correct, and reminding them what we need}
\metaphor{P19}{Terrifyingly effective instant gratification: you immediately have a working thing, but get no benefit from it and you lose your brain}
\metaphor{P22}{``I felt completely dependent on it[AI], and that it was my best friend and greatest enemy''}
\metaphor{P24}{Good at small tasks, but always has to be monitored}
\metaphor{P25}{A coworker: like a code session with somebody more experienced, who you can bounce ideas off of, or who corrects you}
\metaphor{P26}{With great power comes great responsibility}
\metaphor{P29}{A double-edged sword: sometimes it saved me time, other times I felt like I was wasting my time}
\metaphor{P32}{A gateway, giving quicker access to information; and a bridge that you're crossing in a 10-ton truck without knowing the bridge's capacity}
\metaphor{P36}{A junior developer and research assistant that needs a lot of hand-holding}
\metaphor{P37}{My tutor, teacher, expert, architect, programmer, brainstorm partner; not something to rely on solely, but it sped up the process}
\metaphor{P38}{A mentor, because ``he'' knew more than me}
\metaphor{P41}{``like milk and a cup of tea. A little bit would enhance the drink, but too much could ruin it and require you to remake it''}
\metaphor{P42}{Helpful, and manual labor: it took away some of the busy work, filling in the things you didn't really want to do}
\metaphor{P44}{Two steps forward, one step back: part of why I could do anything in this project, but I haven't gained as much as I could; and swimming with water wings for the first time: they make it difficult to swim, can be dangerous when you flip over, and using it feels like a bit of a disservice}
\metaphor*{P45}{A junior developer, because I was managing what code I wanted; but a junior developer that knows everything}
\metaphor{FG46}{A subject expert, because it just knows everything and you can ask it everything}
\metaphor{FG47}{Hasty: it accelerates development without regard for quality; and high level: it let us focus on the broader design while it did the fine things}
\metaphor{FG48}{A time saver, and a funnel: give it something broad and it narrows it down to what you want}
\metaphor{FG50}{An actual assistant, like an unpaid intern}
\metaphor*{FG51}{An ``intelligent gaslighter'': ``I'm right you're wrong trust me on this it should work the way I tell you to''}
\metaphor{FG52}{A guiding assistant: it had more knowledge of the content than me, but it was still an assistant}
\metaphor*{FG53}{``that friend that you meet and think he's really smart [...] you realize he doesn't actually know what he's talking about. And he's been manipulating you''}

%% file: sections/7-limitations.tex
\section{Limitations}
\label{sec:limitations}

Reflexive thematic analysis does not look for the one true account in a dataset. Instead, we used it to build patterns of meaning from data collected in one context~\cite{Braun:2022}. Our themes are therefore not statistically generalizable, and \citet{Braun:2022} do not treat that as a limitation. They ask instead that we explain what may have shaped our reading, so readers can judge whether our findings apply to their own context~\cite{Nowell:2017}. \Cref{sec:trustworthiness} reports how we established trustworthiness during the analysis, and \Cref{sec:discussion} says what our theoretical lens gains and loses. In this section we discuss the setting of this study, and what this setting might have obscured.

We collected data from a single course, and the context of that course will have shaped what participants told us. The course expected AI use, participants knew that and handled accordingly. The conditions also changed mid-project: the teaching team relaxed the deadlines and taught a lecture and a lab on working with AI. Those changes, and their timing, are part of what participants describe, and how they came to reclaim judgment.

We also heard from only some of those teams, and only after completion of the project. Eight teams took part across four sessions, selected from the eleven that expressed interest. We excluded teams that reported interpersonal conflict or non-contributing members, to avoid re-triggering that conflict (\Cref{sec:methodology}). Our accounts therefore come from teams that stayed intact, and we cannot say how teams whose collaboration broke down experienced AI. Participants also spoke in front of their own teammates and a second team, and that will have shaped what they said. The sessions ran after the project ended, so participants worked from memory. We opened the script with a concrete recollection exercise to reduce this effect~\cite{Flick:2000, Stewart:2015}, but what we have are recollections.

Our data are what participants told us. Our analysis is therefore an account of how they made sense of the project. The two can differ: developers' perceptions of their own AI-assisted development do not always match their measured behavior~\cite{Butler:2025, METR:2025:Productivity}. This matters most for our second theme. Where participants described understanding AI-generated code, we can only report that they said so. \citet{James:2024} show that students using AI can be overconfident about their own understanding, and because our data consists of recollections we can not verify their actual understanding.

Another analyst could read our data differently, and \citet{Braun:2022} consider that a property of reflextive thematic analysis, and not a flaw. One alternative reading matters for our first theme. The project was large and open-ended, and novices struggle with such tasks even when they have the skills~\cite{Schoenfeld:2016}. What we read as a surrender to AI may therefore be a response to the task's openness rather than to AI itself. We do not think this replaces our reading, since participants tie the two together and describe reaching for AI when the task overwhelmed them. But the openness of the project and the expectation to use AI were connected, and we cannot separate them.

Our two contrasting themes invite a reader to sort participants into two groups, but it's important to note that participants do not sort that cleanly. Looking for patterns across a dataset can lose the contradictions within one person's account~\cite{Braun:2022}. The same participant could describe surrendering judgment and reclaiming it, and there are instances in our data of this happening (\Cref{sec:findings}). P37 is the clearest case: they regained a sense of capability while both their reliance on AI and the pressure they worked under continued. These instances are important, as they show how closely connected these two experiences were, and how some participants experienced both.

In reflexive thematic analysis, the researcher's subjectivity is a resource rather than a source of bias to be controlled~\cite{Braun:2022}. The first author, who coded the data, teaches software engineering and has worked on software architecture in industry (\Cref{sec:methodology}). This likely made them alert to the loss of understanding participants described, and to what it means to care about a system's quality. It may also have made them quicker to read a hollowed-out account as a loss, and slower to see handing work to AI as a reasonable response to a demanding project. The second author facilitated the focus groups; their experience as a qualitative researcher in psychology and the social sciences helped them probe for detail. During the analysis, they and the first author discussed the coded excerpts to test alternative explanations (\Cref{sec:data-analysis}). 

Through a factual description of the course, the stack, the deadlines, and the expectation to use AI (\Cref{sec:course-context}), we hope to help readers judge whether our account applies to their own setting.

%% file: sections/8-conclusion.tex
\section{Conclusion}
\label{sec:conclusion}

In this study we asked how senior software engineering students \draft{6}{experience AI as they learn software architecture in a team project}. Architecture is a higher-order skill that novices master only through sustained practice, and prior work has focused little on what it is like for students to work with AI while learning it. We studied 43 students on a semester-long team project, and analyzed their accounts through reflexive thematic analysis~\cite{Braun:2022}.

We constructed two themes. \draft{4}{The first, \textbf{Surrendering Judgment to AI}, is an account of how participants spoke about their extensive and deepening reliance on AI, which we interpret as participants eventually surrendering their judgment to it. Participants who felt unprepared ceded decision-making to AI, and that surrender only increased their reliance on AI. They stopped verifying code they could not evaluate, became stuck in a codebase they did not understand, and answered being stuck with more AI rather than with understanding. Prior work has established that novices over-rely on AI and trust it blindly~\cite{Choudhuri:2025:Trenches}. We contribute an account of what that over-reliance was like for the students themselves: how participants ceded decision-making to AI, and the emotional toll they described afterwards, which we have not seen reported before. Participants spoke of shame and frustration, of feeling emotionally disconnected from their work, and of work they called pointless and empty, and we read these accounts as leaving them with even less reason to break the cycle.} Read through Self-Determination Theory~\cite{Ryan:2000}, \draft{4}{these accounts suggest how a reliance on AI can come to erode the motivation novices need to keep learning}.

Our second theme, \draft{3}{\textbf{Applying Human Judgment}}, is an account of \draft{5}{participants who never surrendered their judgment to AI, and of participants who reclaimed it from AI.} It centers on the knowledge and experience that their judgment rested on. It shows that surrender is not inevitable. What separates the themes is not the pressure, which participants worked under in both, but whether participants held the knowledge to judge what AI produced. Some brought that knowledge into the project. Others \draft{3}{recognized that they lacked it, and acquired it} once the course team relaxed the deadlines and taught a lecture and a lab on working with AI, and they then changed how they worked with AI and kept decisions in their own hands.

To us, these first-hand accounts of what surrendering judgment and applying it felt like to participants, and of how some participants moved from one to the other, are an important contribution. \draft{5}{Judgment is what these participants surrendered, and we suspect it is what other developers surrender to AI as well. Studies in other settings, including with experienced developers, could test that.} The conditions that pushed participants to surrender, an unfamiliar stack, delivery pressure outside their control, and an expectation to use AI, are conditions junior developers also meet in industry~\cite{Xin:2024, Begel:2008, Miller:2025}. That is why we ask what is needed to help developers and organizations adopt AI sustainably. Current proposals focus on competence~\cite{Kam:2025}, and our findings point to something more attainable: the participants who recovered did so once the conditions around them changed, and those conditions are controlled by educators and employers.

However, to know when to change the environment, we need the ability to notice when AI adoption goes wrong. Existing instruments do not capture the nuances and subtleties of the accounts we report. And we believe that the words participants gave us, a ``massive crutch'', an ``intelligent gaslighter'', and four months of ``pointless'' prompting, can be used to diagnose such issues. With diagnostics that capture these experiences, we think the judgment, ownership, and motivation of the next generation of developers can be preserved.

%% file: architecture_judgment.bbl

\begin{thebibliography}{70}

%%% the Web address.

\ifx \showCODEN    \undefined \def \showCODEN     #1{\unskip}     \fi
\ifx \showISBNx    \undefined \def \showISBNx     #1{\unskip}     \fi
\ifx \showISBNxiii \undefined \def \showISBNxiii  #1{\unskip}     \fi
\ifx \showISSN     \undefined \def \showISSN      #1{\unskip}     \fi
\ifx \showLCCN     \undefined \def \showLCCN      #1{\unskip}     \fi
\ifx \shownote     \undefined \def \shownote      #1{#1}          \fi
\ifx \showarticletitle \undefined \def \showarticletitle #1{#1}   \fi
\ifx \showURL      \undefined \def \showURL       {\relax}        \fi
% invisible to TeX
\providecommand\bibfield[2]{#2}
\providecommand\bibinfo[2]{#2}
\providecommand\natexlab[1]{#1}
\providecommand\showeprint[2][]{arXiv:#2}

\bibitem[Abbas et~al\mbox{.}(2024)]%
        {Abbas:2024}
\bibfield{author}{\bibinfo{person}{Muhammad Abbas},
  \bibinfo{person}{Farooq~Ahmed Jam}, {and} \bibinfo{person}{Tariq~Iqbal
  Khan}.} \bibinfo{year}{2024}\natexlab{}.
\newblock \showarticletitle{Is it harmful or helpful? Examining the causes and
  consequences of generative AI usage among university students}.
\newblock \bibinfo{journal}{\emph{International Journal of Educational
  Technology in Higher Education}}  \bibinfo{volume}{21} (\bibinfo{date}{2}
  \bibinfo{year}{2024}), \bibinfo{pages}{10}.
\newblock
Issue 1.
\showISSN{2365-9440}
\href{https://doi.org/10.1186/s41239-024-00444-7}{doi:\nolinkurl{10.1186/s41239-024-00444-7}}


\bibitem[Ahmed et~al\mbox{.}(2025)]%
        {Ahmed:2025}
\bibfield{author}{\bibinfo{person}{Sirwan~Khalid Ahmed},
  \bibinfo{person}{Ribwar~Arsalan Mohammed}, \bibinfo{person}{Abdulqadir~J.
  Nashwan}, \bibinfo{person}{Radhwan~Hussein Ibrahim},
  \bibinfo{person}{Araz~Qadir Abdalla}, \bibinfo{person}{Barzan Mohammed~M.
  Ameen}, {and} \bibinfo{person}{Renas~Mohammed Khdhir}.}
  \bibinfo{year}{2025}\natexlab{}.
\newblock \showarticletitle{Using thematic analysis in qualitative research}.
\newblock \bibinfo{journal}{\emph{Journal of Medicine, Surgery, and Public
  Health}}  \bibinfo{volume}{6} (\bibinfo{date}{8} \bibinfo{year}{2025}),
  \bibinfo{pages}{100198}.
\newblock
\showISSN{2949916X}
\href{https://doi.org/10.1016/j.glmedi.2025.100198}{doi:\nolinkurl{10.1016/j.glmedi.2025.100198}}


\bibitem[Alakmeh et~al\mbox{.}(2026)]%
        {Tarek:2026}
\bibfield{author}{\bibinfo{person}{Tarek Alakmeh}, \bibinfo{person}{Norman
  Anderson}, \bibinfo{person}{Victoria Jackson}, \bibinfo{person}{Guilherme~Vaz
  Pereira}, \bibinfo{person}{Umit Akirmak}, \bibinfo{person}{Anthony Estey},
  \bibinfo{person}{Rafael Prikladnicki}, \bibinfo{person}{Andr{\'e} van~der
  Hoek}, \bibinfo{person}{Margaret-Anne Storey}, {and} \bibinfo{person}{Thomas
  Fritz}.} \bibinfo{year}{2026}\natexlab{}.
\newblock \showarticletitle{Grasping AI Reliance in Program Comprehension and
  Coding through the AIRELI Persona Taxonomy}. In
  \bibinfo{booktitle}{\emph{Proceedings of the 2026 34th IEEE/ACM International
  Conference on Program Comprehension}} (Rio de Janeiro, Brazil)
  \emph{(\bibinfo{series}{ICPC '26})}. \bibinfo{publisher}{Association for
  Computing Machinery}, \bibinfo{address}{New York, NY, USA},
  \bibinfo{pages}{123–134}.
\newblock
\showISBNx{9798400724824}
\href{https://doi.org/10.1145/3794763.3794804}{doi:\nolinkurl{10.1145/3794763.3794804}}


\bibitem[Anderson et~al\mbox{.}(2026)]%
        {Anderson:2026:Stuck}
\bibfield{author}{\bibinfo{person}{Norman Anderson}, \bibinfo{person}{Tarek
  Alakmeh}, \bibinfo{person}{Victoria Jackson}, \bibinfo{person}{Guilherme~Vaz
  Pereira}, \bibinfo{person}{Umit Akirmak}, \bibinfo{person}{Anthony Estey},
  \bibinfo{person}{Rafael Prikladnicki}, \bibinfo{person}{Thomas Fritz},
  \bibinfo{person}{André van~der Hoek}, {and} \bibinfo{person}{Margaret-Anne
  Storey}.} \bibinfo{year}{2026}\natexlab{}.
\newblock \bibinfo{title}{ChatGPT: Friend or Foe When Comprehending and
  Changing Unfamiliar Code}.
\newblock
\showeprint[arxiv]{2605.10702}~[cs.SE]
\urldef\tempurl%
\url{https://arxiv.org/abs/2605.10702}
\showURL{%
\tempurl}


\bibitem[Bandura(1998)]%
        {Bandura:1997}
\bibfield{author}{\bibinfo{person}{Albert. Bandura}.}
  \bibinfo{year}{1998}\natexlab{}.
\newblock \bibinfo{booktitle}{\emph{Self-efficacy: the Exercise of Control}}.
\newblock \bibinfo{publisher}{W.H. Freeman}. 604 pages.
\newblock
\showISBNx{9780716728504}


\bibitem[Baresi et~al\mbox{.}(2025)]%
        {Baresi:2025}
\bibfield{author}{\bibinfo{person}{Luciano Baresi}, \bibinfo{person}{Andrea~De
  Lucia}, \bibinfo{person}{Antinisca~Di Marco},
  \bibinfo{person}{Massimiliano~Di Penta}, \bibinfo{person}{Davide~Di Ruscio},
  \bibinfo{person}{Leonardo Mariani}, \bibinfo{person}{Daniela Micucci},
  \bibinfo{person}{Fabio Palomba}, \bibinfo{person}{Maria~Teresa Rossi}, {and}
  \bibinfo{person}{Fiorella Zampetti}.} \bibinfo{year}{2025}\natexlab{}.
\newblock \showarticletitle{Students' Perception of ChatGPT in Software
  Engineering: Lessons Learned from Five Courses}. In
  \bibinfo{booktitle}{\emph{2025 IEEE/ACM 37th International Conference on
  Software Engineering Education and Training (CSEE\&T)}}.
  \bibinfo{publisher}{IEEE}, \bibinfo{pages}{158--169}.
\newblock
\showISBNx{979-8-3315-3709-8}
\href{https://doi.org/10.1109/CSEET66350.2025.00023}{doi:\nolinkurl{10.1109/CSEET66350.2025.00023}}


\bibitem[Barros et~al\mbox{.}(2025)]%
        {Barros:2025}
\bibfield{author}{\bibinfo{person}{Cauã~Ferreira Barros},
  \bibinfo{person}{Bruna~Borges Azevedo}, \bibinfo{person}{Valdemar
  Vicente~Graciano Neto}, \bibinfo{person}{Mohamad Kassab},
  \bibinfo{person}{Marcos Kalinowski}, \bibinfo{person}{Hugo Alexandre D.~Do
  Nascimento}, {and} \bibinfo{person}{Michelle~C.G.S.P. Bandeira}.}
  \bibinfo{year}{2025}\natexlab{}.
\newblock \showarticletitle{Large Language Model for Qualitative Research: A
  Systematic Mapping Study}. In \bibinfo{booktitle}{\emph{2025 IEEE/ACM
  International Workshop on Methodological Issues with Empirical Studies in
  Software Engineering (WSESE)}}. \bibinfo{publisher}{IEEE},
  \bibinfo{pages}{48--55}.
\newblock
\showISBNx{979-8-3315-0225-6}
\href{https://doi.org/10.1109/WSESE66602.2025.00015}{doi:\nolinkurl{10.1109/WSESE66602.2025.00015}}


\bibitem[Bass et~al\mbox{.}(2021)]%
        {Bass:2021}
\bibfield{author}{\bibinfo{person}{Len Bass}, \bibinfo{person}{Paul Clements},
  {and} \bibinfo{person}{Rick Kazman}.} \bibinfo{year}{2021}\natexlab{}.
\newblock \bibinfo{booktitle}{\emph{Software Architecture in Practice}
  (\bibinfo{edition}{4th} ed.)}.
\newblock
\showISBNx{0136886094}


\bibitem[Becker et~al\mbox{.}(2025)]%
        {METR:2025:Productivity}
\bibfield{author}{\bibinfo{person}{Joel Becker}, \bibinfo{person}{Nate Rush},
  \bibinfo{person}{Beth Barnes}, {and} \bibinfo{person}{David Rein}.}
  \bibinfo{year}{2025}\natexlab{}.
\newblock \bibinfo{title}{Measuring the Impact of Early-2025 AI on Experienced
  Open-Source Developer Productivity}.
\newblock
  \bibinfo{howpublished}{\url{https://metr.org/blog/2025-07-10-early-2025-ai-experienced-os-dev-study/}}.
\newblock


\bibitem[Begel and Simon(2008)]%
        {Begel:2008}
\bibfield{author}{\bibinfo{person}{Andrew Begel} {and} \bibinfo{person}{Beth
  Simon}.} \bibinfo{year}{2008}\natexlab{}.
\newblock \showarticletitle{Novice software developers, all over again}. In
  \bibinfo{booktitle}{\emph{Proceedings of the Fourth international Workshop on
  Computing Education Research}} (New York, NY, USA). \bibinfo{publisher}{ACM},
  \bibinfo{pages}{3--14}.
\newblock
\showISBNx{9781605582160}
\href{https://doi.org/10.1145/1404520.1404522}{doi:\nolinkurl{10.1145/1404520.1404522}}


\bibitem[Braun and Clarke(2013)]%
        {Braun:2013}
\bibfield{author}{\bibinfo{person}{Virginia. Braun} {and}
  \bibinfo{person}{Victoria. Clarke}.} \bibinfo{year}{2013}\natexlab{}.
\newblock \bibinfo{booktitle}{\emph{Successful qualitative research: a
  practical guide for beginners}}.
\newblock \bibinfo{publisher}{SAGE Publications Ltd}. 382 pages.
\newblock
\showISBNx{9781847875815}


\bibitem[Braun and Clarke(2022)]%
        {Braun:2022}
\bibfield{author}{\bibinfo{person}{Virginia Braun} {and}
  \bibinfo{person}{Victoria Clarke}.} \bibinfo{year}{2022}\natexlab{}.
\newblock \bibinfo{booktitle}{\emph{Thematic Analysis: a practical guide}}.
\newblock \bibinfo{publisher}{SAGE}. 338 pages.
\newblock
\showISBNx{9781526417305}


\bibitem[Butler et~al\mbox{.}(2025)]%
        {Butler:2025}
\bibfield{author}{\bibinfo{person}{Jenna Butler}, \bibinfo{person}{Jina Suh},
  \bibinfo{person}{Sankeerti Haniyur}, {and} \bibinfo{person}{Constance
  Hadley}.} \bibinfo{year}{2025}\natexlab{}.
\newblock \showarticletitle{Dear Diary: A Randomized Controlled Trial of
  Generative AI Coding Tools in the Workplace}. In
  \bibinfo{booktitle}{\emph{2025 IEEE/ACM 47th International Conference on
  Software Engineering: Software Engineering in Practice (ICSE-SEIP)}}.
  \bibinfo{publisher}{IEEE}, \bibinfo{pages}{319--329}.
\newblock
\showISBNx{979-8-3315-3685-5}
\href{https://doi.org/10.1109/ICSE-SEIP66354.2025.00034}{doi:\nolinkurl{10.1109/ICSE-SEIP66354.2025.00034}}


\bibitem[Choudhuri et~al\mbox{.}(2024)]%
        {Choudhuri:2024}
\bibfield{author}{\bibinfo{person}{Rudrajit Choudhuri}, \bibinfo{person}{Dylan
  Liu}, \bibinfo{person}{Igor Steinmacher}, \bibinfo{person}{Marco Gerosa},
  {and} \bibinfo{person}{Anita Sarma}.} \bibinfo{year}{2024}\natexlab{}.
\newblock \showarticletitle{How Far Are We? The Triumphs and Trials of
  Generative AI in Learning Software Engineering}. In
  \bibinfo{booktitle}{\emph{Proceedings of the IEEE/ACM 46th International
  Conference on Software Engineering}} (New York, NY, USA).
  \bibinfo{publisher}{ACM}, \bibinfo{pages}{1--13}.
\newblock
\showISBNx{9798400702174}
\href{https://doi.org/10.1145/3597503.3639201}{doi:\nolinkurl{10.1145/3597503.3639201}}


\bibitem[Choudhuri et~al\mbox{.}(2025)]%
        {Choudhuri:2025:Trenches}
\bibfield{author}{\bibinfo{person}{Rudrajit Choudhuri},
  \bibinfo{person}{Ambareesh Ramakrishnan}, \bibinfo{person}{Amreeta
  Chatterjee}, \bibinfo{person}{Bianca Trinkenreich}, \bibinfo{person}{Igor
  Steinmacher}, \bibinfo{person}{Marco Gerosa}, {and} \bibinfo{person}{Anita
  Sarma}.} \bibinfo{year}{2025}\natexlab{}.
\newblock \showarticletitle{Insights from the Frontline: GenAI Utilization
  Among Software Engineering Students}. In \bibinfo{booktitle}{\emph{2025
  IEEE/ACM 37th International Conference on Software Engineering Education and
  Training (CSEE\&T)}}. \bibinfo{publisher}{IEEE}, \bibinfo{pages}{1--12}.
\newblock
\showISBNx{979-8-3315-3709-8}
\href{https://doi.org/10.1109/CSEET66350.2025.00007}{doi:\nolinkurl{10.1109/CSEET66350.2025.00007}}


\bibitem[Choudhuri et~al\mbox{.}(2026)]%
        {Choudhuri:2026}
\bibfield{author}{\bibinfo{person}{Rudrajit Choudhuri},
  \bibinfo{person}{Christopher Sanchez}, \bibinfo{person}{Margaret Burnett},
  {and} \bibinfo{person}{Anita Sarma}.} \bibinfo{year}{2026}\natexlab{}.
\newblock \bibinfo{title}{Thinking Less, Trusting More: GenAI's Impacts on
  Students' Cognitive Habits}.
\newblock
\showeprint[arxiv]{2601.22430}~[cs.HC]
\urldef\tempurl%
\url{https://arxiv.org/abs/2601.22430}
\showURL{%
\tempurl}


\bibitem[Csikszentmihalyi(1990)]%
        {Csikszentmihalyi:1990}
\bibfield{author}{\bibinfo{person}{Mihaly Csikszentmihalyi}.}
  \bibinfo{year}{1990}\natexlab{}.
\newblock \bibinfo{booktitle}{\emph{Flow: The Psychology of Optimal
  Experience}}.
\newblock \bibinfo{publisher}{Harper \& Row}, \bibinfo{address}{New York}.
\newblock
\showISBNx{0060920432}


\bibitem[Deursen et~al\mbox{.}(2017)]%
        {Deursen:2017}
\bibfield{author}{\bibinfo{person}{Arie~Van Deursen},
  \bibinfo{person}{Maurício Aniche}, \bibinfo{person}{Joop Aué},
  \bibinfo{person}{Rogier Slag}, \bibinfo{person}{Michael~De Jong},
  \bibinfo{person}{Alex Nederlof}, {and} \bibinfo{person}{Eric Bouwers}.}
  \bibinfo{year}{2017}\natexlab{}.
\newblock \showarticletitle{A Collaborative Approach to Teaching Software
  Architecture}. In \bibinfo{booktitle}{\emph{Proceedings of the 2017 ACM
  SIGCSE Technical Symposium on Computer Science Education}} (New York, NY,
  USA). \bibinfo{publisher}{ACM}, \bibinfo{pages}{591--596}.
\newblock
\showISBNx{9781450346986}
\href{https://doi.org/10.1145/3017680.3017737}{doi:\nolinkurl{10.1145/3017680.3017737}}


\bibitem[Ernst and Treude(2026)]%
        {Ernst:2026}
\bibfield{author}{\bibinfo{person}{Neil~A. Ernst} {and}
  \bibinfo{person}{Christoph Treude}.} \bibinfo{year}{2026}\natexlab{}.
\newblock \showarticletitle{GenAI Is No Silver Bullet for Qualitative Research
  in Software Engineering}.
\newblock \bibinfo{journal}{\emph{ACM Transactions on Software Engineering and
  Methodology}} (\bibinfo{date}{7} \bibinfo{year}{2026}).
\newblock
\showISSN{1049-331X}
\href{https://doi.org/10.1145/3837087}{doi:\nolinkurl{10.1145/3837087}}


\bibitem[Esposito et~al\mbox{.}(2026)]%
        {Esposito:2026}
\bibfield{author}{\bibinfo{person}{Matteo Esposito}, \bibinfo{person}{Xiaozhou
  Li}, \bibinfo{person}{Sergio Moreschini}, \bibinfo{person}{Noman Ahmad},
  \bibinfo{person}{Tomas Cerny}, \bibinfo{person}{Karthik Vaidhyanathan},
  \bibinfo{person}{Valentina Lenarduzzi}, {and} \bibinfo{person}{Davide
  Taibi}.} \bibinfo{year}{2026}\natexlab{}.
\newblock \showarticletitle{Generative AI for software architecture.
  Applications, challenges, and future directions}.
\newblock \bibinfo{journal}{\emph{Journal of Systems and Software}}
  \bibinfo{volume}{231} (\bibinfo{date}{1} \bibinfo{year}{2026}),
  \bibinfo{pages}{112607}.
\newblock
\showISSN{01641212}
\href{https://doi.org/10.1016/j.jss.2025.112607}{doi:\nolinkurl{10.1016/j.jss.2025.112607}}


\bibitem[Fenu et~al\mbox{.}(2024)]%
        {Fenu:2024}
\bibfield{author}{\bibinfo{person}{Gianni Fenu}, \bibinfo{person}{Roberta
  Galici}, \bibinfo{person}{Mirko Marras}, {and} \bibinfo{person}{Diego
  Reforgiato}.} \bibinfo{year}{2024}\natexlab{}.
\newblock \showarticletitle{Exploring Student Interactions with AI in
  Programming Training}. In \bibinfo{booktitle}{\emph{Adjunct Proceedings of
  the 32nd ACM Conference on User Modeling, Adaptation and Personalization}}
  (New York, NY, USA). \bibinfo{publisher}{ACM}, \bibinfo{pages}{555--560}.
\newblock
\showISBNx{9798400704666}
\href{https://doi.org/10.1145/3631700.3665227}{doi:\nolinkurl{10.1145/3631700.3665227}}


\bibitem[Ferino et~al\mbox{.}(2025)]%
        {Ferino:2025:STGT}
\bibfield{author}{\bibinfo{person}{Samuel Ferino}, \bibinfo{person}{Rashina
  Hoda}, \bibinfo{person}{John Grundy}, {and} \bibinfo{person}{Christoph
  Treude}.} \bibinfo{year}{2025}\natexlab{}.
\newblock \bibinfo{title}{Walking the Tightrope of LLMs for Software
  Development: A Practitioners' Perspective}.
\newblock
\showeprint[arxiv]{2511.06428}~[cs.SE]
\urldef\tempurl%
\url{https://arxiv.org/abs/2511.06428}
\showURL{%
\tempurl}


\bibitem[Ferino et~al\mbox{.}(2026)]%
        {Ferino:2026}
\bibfield{author}{\bibinfo{person}{Samuel Ferino}, \bibinfo{person}{Rashina
  Hoda}, \bibinfo{person}{John Grundy}, {and} \bibinfo{person}{Christoph
  Treude}.} \bibinfo{year}{2026}\natexlab{}.
\newblock \showarticletitle{Novice Developers’ Perspectives on Adopting LLMs
  for Software Development: A Systematic Literature Review}.
\newblock \bibinfo{journal}{\emph{ACM Transactions on Software Engineering and
  Methodology}} (\bibinfo{date}{3} \bibinfo{year}{2026}).
\newblock
\showISSN{1049-331X}
\href{https://doi.org/10.1145/3800580}{doi:\nolinkurl{10.1145/3800580}}


\bibitem[Flick(2000)]%
        {Flick:2000}
\bibfield{author}{\bibinfo{person}{Uwe Flick}.}
  \bibinfo{year}{2000}\natexlab{}.
\newblock \bibinfo{booktitle}{\emph{Episodic Interviewing}}.
\newblock \bibinfo{publisher}{SAGE Publications Ltd}, \bibinfo{pages}{76--92}.
\newblock
\href{https://doi.org/10.4135/9781849209731.n5}{doi:\nolinkurl{10.4135/9781849209731.n5}}


\bibitem[Galster and Angelov(2016)]%
        {Galster:2016}
\bibfield{author}{\bibinfo{person}{Matthias Galster} {and}
  \bibinfo{person}{Samuil Angelov}.} \bibinfo{year}{2016}\natexlab{}.
\newblock \showarticletitle{What makes teaching software architecture
  difficult?}. In \bibinfo{booktitle}{\emph{Proceedings of the 38th
  International Conference on Software Engineering Companion}} (New York, NY,
  USA). \bibinfo{publisher}{ACM}, \bibinfo{pages}{356--359}.
\newblock
\showISBNx{9781450342056}
\href{https://doi.org/10.1145/2889160.2889187}{doi:\nolinkurl{10.1145/2889160.2889187}}


\bibitem[Ghimire and Edwards(2024)]%
        {Aashish:2024}
\bibfield{author}{\bibinfo{person}{Aashish Ghimire} {and} \bibinfo{person}{John
  Edwards}.} \bibinfo{year}{2024}\natexlab{}.
\newblock \bibinfo{booktitle}{\emph{Coding with AI: How Are Tools Like ChatGPT
  Being Used by Students in Foundational Programming Courses}}.
\newblock \bibinfo{pages}{259--267}.
\newblock
\href{https://doi.org/10.1007/978-3-031-64299-9_20}{doi:\nolinkurl{10.1007/978-3-031-64299-9_20}}


\bibitem[Groothuijsen et~al\mbox{.}(2024)]%
        {Groothuijsen:2024}
\bibfield{author}{\bibinfo{person}{Suzanne Groothuijsen},
  \bibinfo{person}{Antoine van~den Beemt}, \bibinfo{person}{Joris~C. Remmers},
  {and} \bibinfo{person}{Ludo~W. van Meeuwen}.}
  \bibinfo{year}{2024}\natexlab{}.
\newblock \showarticletitle{AI chatbots in programming education: Students’
  use in a scientific computing course and consequences for learning}.
\newblock \bibinfo{journal}{\emph{Computers and Education: Artificial
  Intelligence}}  \bibinfo{volume}{7} (\bibinfo{date}{12}
  \bibinfo{year}{2024}).
\newblock
\showISSN{2666920X}
\href{https://doi.org/10.1016/j.caeai.2024.100290}{doi:\nolinkurl{10.1016/j.caeai.2024.100290}}


\bibitem[Guerra and Ernst(2025)]%
        {Guerra:2025}
\bibfield{author}{\bibinfo{person}{Luiz~Franciscatto Guerra} {and}
  \bibinfo{person}{Neil Ernst}.} \bibinfo{year}{2025}\natexlab{}.
\newblock \bibinfo{title}{Assessing LLMs for Front-end Software Architecture
  Knowledge}.
\newblock
\showeprint[arxiv]{2502.19518}~[cs.SE]
\urldef\tempurl%
\url{https://arxiv.org/abs/2502.19518}
\showURL{%
\tempurl}


\bibitem[Güner and Er(2025)]%
        {Guner:2025}
\bibfield{author}{\bibinfo{person}{Hacer Güner} {and} \bibinfo{person}{Erkan
  Er}.} \bibinfo{year}{2025}\natexlab{}.
\newblock \showarticletitle{AI in the classroom: Exploring students’
  interaction with ChatGPT in programming learning}.
\newblock \bibinfo{journal}{\emph{Education and Information Technologies}}
  \bibinfo{volume}{30} (\bibinfo{date}{6} \bibinfo{year}{2025}),
  \bibinfo{pages}{12681--12707}.
\newblock
Issue 9.
\showISSN{1360-2357}
\href{https://doi.org/10.1007/s10639-025-13337-7}{doi:\nolinkurl{10.1007/s10639-025-13337-7}}


\bibitem[Haindl and Weinberger(2024)]%
        {Haindl:2024}
\bibfield{author}{\bibinfo{person}{Philipp Haindl} {and}
  \bibinfo{person}{Gerald Weinberger}.} \bibinfo{year}{2024}\natexlab{}.
\newblock \showarticletitle{Students’ Experiences of Using ChatGPT in an
  Undergraduate Programming Course}.
\newblock \bibinfo{journal}{\emph{IEEE Access}}  \bibinfo{volume}{12}
  (\bibinfo{year}{2024}), \bibinfo{pages}{43519--43529}.
\newblock
\showISSN{2169-3536}
\href{https://doi.org/10.1109/ACCESS.2024.3380909}{doi:\nolinkurl{10.1109/ACCESS.2024.3380909}}


\bibitem[Hasselbring(2018)]%
        {Hasselbring:2018}
\bibfield{author}{\bibinfo{person}{Wilhelm Hasselbring}.}
  \bibinfo{year}{2018}\natexlab{}.
\newblock \bibinfo{booktitle}{\emph{Software Architecture: Past, Present,
  Future}}.
\newblock \bibinfo{publisher}{Springer International Publishing},
  \bibinfo{pages}{169--184}.
\newblock
\href{https://doi.org/10.1007/978-3-319-73897-0_10}{doi:\nolinkurl{10.1007/978-3-319-73897-0_10}}


\bibitem[Jaman et~al\mbox{.}(2026)]%
        {Jaman:2026}
\bibfield{author}{\bibinfo{person}{Md~Gulam~Mashud Jaman},
  \bibinfo{person}{Abdul Mouhit}, {and} \bibinfo{person}{Georgios
  Koutsopoulos}.} \bibinfo{year}{2026}\natexlab{}.
\newblock \bibinfo{booktitle}{\emph{Understanding Generative AI Adoption in
  Development Work: The Role of Developer Experience}}.
\newblock \bibinfo{pages}{316--332}.
\newblock
\href{https://doi.org/10.1007/978-3-032-30863-4_20}{doi:\nolinkurl{10.1007/978-3-032-30863-4_20}}


\bibitem[Jowsey et~al\mbox{.}(2025)]%
        {Jowsey:2025}
\bibfield{author}{\bibinfo{person}{Tanisha Jowsey}, \bibinfo{person}{Virginia
  Braun}, \bibinfo{person}{Victoria Clarke}, \bibinfo{person}{Deborah Lupton},
  {and} \bibinfo{person}{Michelle Fine}.} \bibinfo{year}{2025}\natexlab{}.
\newblock \showarticletitle{We Reject the Use of Generative Artificial
  Intelligence for Reflexive Qualitative Research}.
\newblock \bibinfo{journal}{\emph{Qualitative Inquiry}} (\bibinfo{date}{12}
  \bibinfo{year}{2025}).
\newblock
\showISSN{1077-8004}
\href{https://doi.org/10.1177/10778004251401851}{doi:\nolinkurl{10.1177/10778004251401851}}


\bibitem[Kam et~al\mbox{.}(2025)]%
        {Kam:2025}
\bibfield{author}{\bibinfo{person}{Matthew Kam}, \bibinfo{person}{Cody Miller},
  \bibinfo{person}{Miaoxin Wang}, \bibinfo{person}{Abey Tidwell},
  \bibinfo{person}{Irene~A. Lee}, \bibinfo{person}{Joyce Malyn-Smith},
  \bibinfo{person}{Beatriz Perret}, \bibinfo{person}{Vikram Tiwari},
  \bibinfo{person}{Joshua Kenitzer}, \bibinfo{person}{Andrew Macvean}, {and}
  \bibinfo{person}{Erin Barrar}.} \bibinfo{year}{2025}\natexlab{}.
\newblock \showarticletitle{What do professional software developers need to
  know to succeed in an age of Artificial Intelligence?}. In
  \bibinfo{booktitle}{\emph{Proceedings of the 33rd ACM International
  Conference on the Foundations of Software Engineering}} (New York, NY, USA).
  \bibinfo{publisher}{ACM}, \bibinfo{pages}{947--958}.
\newblock
\showISBNx{9798400712760}
\href{https://doi.org/10.1145/3696630.3727251}{doi:\nolinkurl{10.1145/3696630.3727251}}


\bibitem[Keuning et~al\mbox{.}(2024)]%
        {Keuning:2024}
\bibfield{author}{\bibinfo{person}{Hieke Keuning}, \bibinfo{person}{Isaac
  Alpizar-Chacon}, \bibinfo{person}{Ioanna Lykourentzou},
  \bibinfo{person}{Lauren Beehler}, \bibinfo{person}{Christian Köppe},
  \bibinfo{person}{Imke de Jong}, {and} \bibinfo{person}{Sergey Sosnovsky}.}
  \bibinfo{year}{2024}\natexlab{}.
\newblock \showarticletitle{Students' Perceptions and Use of Generative AI
  Tools for Programming Across Different Computing Courses}. In
  \bibinfo{booktitle}{\emph{Proceedings of the 24th Koli Calling International
  Conference on Computing Education Research}} (New York, NY, USA).
  \bibinfo{publisher}{ACM}, \bibinfo{pages}{1--12}.
\newblock
\showISBNx{9798400710384}
\href{https://doi.org/10.1145/3699538.3699546}{doi:\nolinkurl{10.1145/3699538.3699546}}


\bibitem[Kosmyna et~al\mbox{.}(2025)]%
        {Kosmyna:2025}
\bibfield{author}{\bibinfo{person}{Nataliya Kosmyna}, \bibinfo{person}{Eugene
  Hauptmann}, \bibinfo{person}{Ye~Tong Yuan}, \bibinfo{person}{Jessica Situ},
  \bibinfo{person}{Xian-Hao Liao}, \bibinfo{person}{Ashly~Vivian Beresnitzky},
  \bibinfo{person}{Iris Braunstein}, {and} \bibinfo{person}{Pattie Maes}.}
  \bibinfo{year}{2025}\natexlab{}.
\newblock \bibinfo{title}{Your Brain on ChatGPT: Accumulation of Cognitive Debt
  when Using an AI Assistant for Essay Writing Task}.
\newblock
\showeprint[arxiv]{2506.08872}~[cs.AI]
\urldef\tempurl%
\url{https://arxiv.org/abs/2506.08872}
\showURL{%
\tempurl}


\bibitem[Kusurkar et~al\mbox{.}(2013)]%
        {Kusurkar:2013}
\bibfield{author}{\bibinfo{person}{R.~A. Kusurkar}, \bibinfo{person}{Th. J.~Ten
  Cate}, \bibinfo{person}{C.~M.~P. Vos}, \bibinfo{person}{P. Westers}, {and}
  \bibinfo{person}{G. Croiset}.} \bibinfo{year}{2013}\natexlab{}.
\newblock \showarticletitle{How motivation affects academic performance: a
  structural equation modelling analysis}.
\newblock \bibinfo{journal}{\emph{Advances in Health Sciences Education}}
  \bibinfo{volume}{18} (\bibinfo{date}{3} \bibinfo{year}{2013}),
  \bibinfo{pages}{57--69}.
\newblock
Issue 1.
\showISSN{1382-4996}
\href{https://doi.org/10.1007/s10459-012-9354-3}{doi:\nolinkurl{10.1007/s10459-012-9354-3}}


\bibitem[Loksa et~al\mbox{.}(2016)]%
        {Loksa:2016}
\bibfield{author}{\bibinfo{person}{Dastyni Loksa}, \bibinfo{person}{Amy~J. Ko},
  \bibinfo{person}{Will Jernigan}, \bibinfo{person}{Alannah Oleson},
  \bibinfo{person}{Christopher~J. Mendez}, {and} \bibinfo{person}{Margaret~M.
  Burnett}.} \bibinfo{year}{2016}\natexlab{}.
\newblock \showarticletitle{Programming, Problem Solving, and Self-Awareness}.
  In \bibinfo{booktitle}{\emph{Proceedings of the 2016 CHI Conference on Human
  Factors in Computing Systems}} (New York, NY, USA). \bibinfo{publisher}{ACM},
  \bibinfo{pages}{1449--1461}.
\newblock
\showISBNx{9781450333627}
\href{https://doi.org/10.1145/2858036.2858252}{doi:\nolinkurl{10.1145/2858036.2858252}}


\bibitem[Margulieux et~al\mbox{.}(2024)]%
        {Margulieux:2024}
\bibfield{author}{\bibinfo{person}{Lauren~E. Margulieux},
  \bibinfo{person}{James Prather}, \bibinfo{person}{Brent~N. Reeves},
  \bibinfo{person}{Brett~A. Becker}, \bibinfo{person}{Gozde~Cetin Uzun},
  \bibinfo{person}{Dastyni Loksa}, \bibinfo{person}{Juho Leinonen}, {and}
  \bibinfo{person}{Paul Denny}.} \bibinfo{year}{2024}\natexlab{}.
\newblock \showarticletitle{Self-Regulation, Self-Efficacy, and Fear of Failure
  Interactions with How Novices Use LLMs to Solve Programming Problems}. In
  \bibinfo{booktitle}{\emph{Proceedings of the 2024 on Innovation and
  Technology in Computer Science Education V. 1}} (New York, NY, USA).
  \bibinfo{publisher}{ACM}, \bibinfo{pages}{276--282}.
\newblock
\showISBNx{9798400706004}
\href{https://doi.org/10.1145/3649217.3653621}{doi:\nolinkurl{10.1145/3649217.3653621}}


\bibitem[Miles(2026)]%
        {Miles:2026}
\bibfield{author}{\bibinfo{person}{Russ Miles}.}
  \bibinfo{year}{2026}\natexlab{}.
\newblock \bibinfo{booktitle}{\emph{The Sovereign Apprentice: AI Literacy for
  New Engineers}}.
\newblock \bibinfo{publisher}{Leanpub}.
\newblock
\urldef\tempurl%
\url{https://leanpub.com/thesovereignapprentice}
\showURL{%
\tempurl}


\bibitem[Miller et~al\mbox{.}(2025)]%
        {Miller:2025}
\bibfield{author}{\bibinfo{person}{Courtney Miller}, \bibinfo{person}{Rudrajit
  Choudhuri}, \bibinfo{person}{Mara Ulloa}, \bibinfo{person}{Sankeerti
  Haniyur}, \bibinfo{person}{Robert DeLine}, \bibinfo{person}{Margaret-Anne
  Storey}, \bibinfo{person}{Emerson Murphy-Hill}, \bibinfo{person}{Christian
  Bird}, {and} \bibinfo{person}{Jenna~L. Butler}.}
  \bibinfo{year}{2025}\natexlab{}.
\newblock \bibinfo{title}{"Maybe We Need Some More Examples:" Individual and
  Team Drivers of Developer GenAI Tool Use}.
\newblock
\showeprint[arxiv]{2507.21280}~[cs.SE]


\bibitem[Mircea et~al\mbox{.}(2026)]%
        {Mircea:2026}
\bibfield{author}{\bibinfo{person}{Michael Mircea}, \bibinfo{person}{Elisa
  Schmid}, \bibinfo{person}{Jakob Droste}, {and} \bibinfo{person}{Kurt
  Schneider}.} \bibinfo{year}{2026}\natexlab{}.
\newblock \bibinfo{title}{How Do Software Engineering Students Use Generative
  AI in Real-World Capstone Projects? An Empirical Baseline Study}.
\newblock
\showeprint[arxiv]{2604.24521}~[cs.SE]
\urldef\tempurl%
\url{https://arxiv.org/abs/2604.24521}
\showURL{%
\tempurl}


\bibitem[Mitas and Sachweh(2025)]%
        {Mitas:2025}
\bibfield{author}{\bibinfo{person}{Marcel Mitas} {and} \bibinfo{person}{Sabine
  Sachweh}.} \bibinfo{year}{2025}\natexlab{}.
\newblock \showarticletitle{Lifelong Learning in Software Engineering: Towards
  an AI-driven Tutor for Cloud-Based Software Architectures}. In
  \bibinfo{booktitle}{\emph{2025 IEEE European Technology and Engineering
  Management Summit (E-TEMS)}}. \bibinfo{publisher}{IEEE},
  \bibinfo{pages}{92--97}.
\newblock
\showISBNx{979-8-3315-1150-0}
\href{https://doi.org/10.1109/E-TEMS64751.2025.11239087}{doi:\nolinkurl{10.1109/E-TEMS64751.2025.11239087}}


\bibitem[Mohammed and Rocke(2025)]%
        {Mohammed:2025}
\bibfield{author}{\bibinfo{person}{Crista Mohammed} {and} \bibinfo{person}{Sean
  Rocke}.} \bibinfo{year}{2025}\natexlab{}.
\newblock \showarticletitle{Electrical and Computer Engineering Freshmen and
  Generative AI: Awareness, Attitudes, and Ethics}. In
  \bibinfo{booktitle}{\emph{2025 IEEE Global Engineering Education Conference
  (EDUCON)}}. \bibinfo{publisher}{IEEE}, \bibinfo{pages}{1--8}.
\newblock
\showISBNx{979-8-3315-3949-8}
\href{https://doi.org/10.1109/EDUCON62633.2025.11016545}{doi:\nolinkurl{10.1109/EDUCON62633.2025.11016545}}


\bibitem[Montes et~al\mbox{.}(2025)]%
        {Montes:2025}
\bibfield{author}{\bibinfo{person}{Cristina~Martinez Montes},
  \bibinfo{person}{Robert Feldt}, \bibinfo{person}{Cristina~Miguel Martos},
  \bibinfo{person}{Sofia Ouhbi}, \bibinfo{person}{Shweta Premanandan}, {and}
  \bibinfo{person}{Daniel Graziotin}.} \bibinfo{year}{2025}\natexlab{}.
\newblock \bibinfo{title}{Large Language Models in Thematic Analysis: Prompt
  Engineering, Evaluation, and Guidelines for Qualitative Software Engineering
  Research}.
\newblock
\showeprint[arxiv]{2510.18456}~[cs.SE]
\urldef\tempurl%
\url{https://arxiv.org/abs/2510.18456}
\showURL{%
\tempurl}


\bibitem[Nowell et~al\mbox{.}(2017)]%
        {Nowell:2017}
\bibfield{author}{\bibinfo{person}{Lorelli~S. Nowell}, \bibinfo{person}{Jill~M.
  Norris}, \bibinfo{person}{Deborah~E. White}, {and} \bibinfo{person}{Nancy~J.
  Moules}.} \bibinfo{year}{2017}\natexlab{}.
\newblock \showarticletitle{Thematic Analysis}.
\newblock \bibinfo{journal}{\emph{International Journal of Qualitative
  Methods}}  \bibinfo{volume}{16} (\bibinfo{date}{12} \bibinfo{year}{2017}).
\newblock
Issue 1.
\showISSN{1609-4069}
\href{https://doi.org/10.1177/1609406917733847}{doi:\nolinkurl{10.1177/1609406917733847}}


\bibitem[Oliveira et~al\mbox{.}(2025)]%
        {Oliveira:2025}
\bibfield{author}{\bibinfo{person}{Eduardo~Carneiro Oliveira},
  \bibinfo{person}{Hieke Keuning}, {and} \bibinfo{person}{Johan Jeuring}.}
  \bibinfo{year}{2025}\natexlab{}.
\newblock \showarticletitle{Uncovering Behavioral Patterns in Student–LLM
  Conversations during Code Refactoring Tasks}. In
  \bibinfo{booktitle}{\emph{Proceedings of the 25th Koli Calling International
  Conference on Computing Education Research}} (New York, NY, USA).
  \bibinfo{publisher}{ACM}, \bibinfo{pages}{1--11}.
\newblock
\showISBNx{9798400715990}
\href{https://doi.org/10.1145/3769994.3770000}{doi:\nolinkurl{10.1145/3769994.3770000}}


\bibitem[Pearson et~al\mbox{.}(2025)]%
        {Pearson:2025}
\bibfield{author}{\bibinfo{person}{Helen Pearson}, \bibinfo{person}{Michelle
  Myall}, \bibinfo{person}{Anne-Sophie Darlington}, {and}
  \bibinfo{person}{Faith Gibson}.} \bibinfo{year}{2025}\natexlab{}.
\newblock \showarticletitle{The approach and application of analysing inductive
  and deductive datasets: a worked example using reflexive thematic analysis}.
\newblock \bibinfo{journal}{\emph{Qualitative Research in Psychology}}
  \bibinfo{volume}{22} (\bibinfo{date}{10} \bibinfo{year}{2025}),
  \bibinfo{pages}{842--886}.
\newblock
Issue 4.
\showISSN{1478-0887}
\href{https://doi.org/10.1080/14780887.2025.2499265}{doi:\nolinkurl{10.1080/14780887.2025.2499265}}


\bibitem[Prather et~al\mbox{.}(2024)]%
        {James:2024}
\bibfield{author}{\bibinfo{person}{James Prather}, \bibinfo{person}{Brent~N
  Reeves}, \bibinfo{person}{Juho Leinonen}, \bibinfo{person}{Stephen MacNeil},
  \bibinfo{person}{Arisoa~S Randrianasolo}, \bibinfo{person}{Brett~A. Becker},
  \bibinfo{person}{Bailey Kimmel}, \bibinfo{person}{Jared Wright}, {and}
  \bibinfo{person}{Ben Briggs}.} \bibinfo{year}{2024}\natexlab{}.
\newblock \showarticletitle{The Widening Gap: The Benefits and Harms of
  Generative AI for Novice Programmers}. In
  \bibinfo{booktitle}{\emph{Proceedings of the 2024 ACM Conference on
  International Computing Education Research - Volume 1}} (New York, NY, USA).
  \bibinfo{publisher}{ACM}, \bibinfo{pages}{469--486}.
\newblock
\showISBNx{9798400704758}
\href{https://doi.org/10.1145/3632620.3671116}{doi:\nolinkurl{10.1145/3632620.3671116}}


\bibitem[Radford et~al\mbox{.}(2023)]%
        {Radford:2023}
\bibfield{author}{\bibinfo{person}{Alec Radford}, \bibinfo{person}{Jong~Wook
  Kim}, \bibinfo{person}{Tao Xu}, \bibinfo{person}{Greg Brockman},
  \bibinfo{person}{Christine McLeavey}, {and} \bibinfo{person}{Ilya
  Sutskever}.} \bibinfo{year}{2023}\natexlab{}.
\newblock \showarticletitle{Robust speech recognition via large-scale weak
  supervision}. In \bibinfo{booktitle}{\emph{Proceedings of the 40th
  International Conference on Machine Learning}} (Honolulu, Hawaii, USA)
  \emph{(\bibinfo{series}{ICML'23})}. \bibinfo{publisher}{JMLR.org}, Article
  \bibinfo{articleno}{1182}, \bibinfo{numpages}{27}~pages.
\newblock


\bibitem[Rahe and Maalej(2025)]%
        {Rahe:2025}
\bibfield{author}{\bibinfo{person}{Christian Rahe} {and} \bibinfo{person}{Walid
  Maalej}.} \bibinfo{year}{2025}\natexlab{}.
\newblock \showarticletitle{How Do Programming Students Use Generative AI?}
\newblock \bibinfo{journal}{\emph{Proceedings of the ACM on Software
  Engineering}}  \bibinfo{volume}{2} (\bibinfo{date}{6} \bibinfo{year}{2025}),
  \bibinfo{pages}{978--1000}.
\newblock
Issue FSE.
\showISSN{2994-970X}
\href{https://doi.org/10.1145/3715762}{doi:\nolinkurl{10.1145/3715762}}


\bibitem[Ramalingam and Wiedenbeck(1998)]%
        {Ramalingam:1998}
\bibfield{author}{\bibinfo{person}{Vennila Ramalingam} {and}
  \bibinfo{person}{Susan Wiedenbeck}.} \bibinfo{year}{1998}\natexlab{}.
\newblock \showarticletitle{Development and Validation of Scores on a Computer
  Programming Self-Efficacy Scale and Group Analyses of Novice Programmer
  Self-Efficacy}.
\newblock \bibinfo{journal}{\emph{Journal of Educational Computing Research}}
  \bibinfo{volume}{19} (\bibinfo{date}{12} \bibinfo{year}{1998}),
  \bibinfo{pages}{367--381}.
\newblock
Issue 4.
\showISSN{0735-6331}
\href{https://doi.org/10.2190/C670-Y3C8-LTJ1-CT3P}{doi:\nolinkurl{10.2190/C670-Y3C8-LTJ1-CT3P}}


\bibitem[Rojas-Galeano et~al\mbox{.}(2025)]%
        {Rojas-Galeano:2025}
\bibfield{author}{\bibinfo{person}{Sergio Rojas-Galeano},
  \bibinfo{person}{Julian Tejada}, {and} \bibinfo{person}{Fernando
  Marmolejo-Ramos}.} \bibinfo{year}{2025}\natexlab{}.
\newblock \bibinfo{title}{Between Tool and Trouble: Student Attitudes Toward AI
  in Programming Education}.
\newblock
\showeprint[arxiv]{2508.05999}~[cs.ET]
\urldef\tempurl%
\url{https://arxiv.org/abs/2508.05999}
\showURL{%
\tempurl}


\bibitem[Runeson et~al\mbox{.}(2012)]%
        {Runeson:2012}
\bibfield{author}{\bibinfo{person}{Per Runeson}, \bibinfo{person}{Martin Host},
  \bibinfo{person}{Austen Rainer}, {and} \bibinfo{person}{Bjorn Regnell}.}
  \bibinfo{year}{2012}\natexlab{}.
\newblock \bibinfo{booktitle}{\emph{Case Study Research in Software
  Engineering}}.
\newblock 241 pages.
\newblock
\showISBNx{9781118104354}


\bibitem[Ryan and Deci(2000)]%
        {Ryan:2000}
\bibfield{author}{\bibinfo{person}{Richard~M. Ryan} {and}
  \bibinfo{person}{Edward~L. Deci}.} \bibinfo{year}{2000}\natexlab{}.
\newblock \showarticletitle{Self-determination theory and the facilitation of
  intrinsic motivation, social development, and well-being.}
\newblock \bibinfo{journal}{\emph{American Psychologist}}  \bibinfo{volume}{55}
  (\bibinfo{year}{2000}), \bibinfo{pages}{68--78}.
\newblock
Issue 1.
\showISSN{1935-990X}
\href{https://doi.org/10.1037/0003-066X.55.1.68}{doi:\nolinkurl{10.1037/0003-066X.55.1.68}}


\bibitem[Salomon et~al\mbox{.}(2025)]%
        {Salomon:2025}
\bibfield{author}{\bibinfo{person}{Marie Salomon}, \bibinfo{person}{Kyle~D.
  Chin}, \bibinfo{person}{Reid Holmes}, \bibinfo{person}{Thomas Fritz}, {and}
  \bibinfo{person}{Gail~C. Murphy}.} \bibinfo{year}{2025}\natexlab{}.
\newblock \showarticletitle{An Exploration of How Generative AI Affects
  Workflow and Collaboration in a Software Engineering Course}. In
  \bibinfo{booktitle}{\emph{Proceedings of the 2025 ACM SIGPLAN International
  Symposium on SPLASH-E}} (New York, NY, USA). \bibinfo{publisher}{ACM},
  \bibinfo{pages}{42--54}.
\newblock
\showISBNx{9798400721427}
\href{https://doi.org/10.1145/3758317.3759680}{doi:\nolinkurl{10.1145/3758317.3759680}}


\bibitem[Schmid et~al\mbox{.}(2025)]%
        {Schmid:2025}
\bibfield{author}{\bibinfo{person}{Larissa Schmid}, \bibinfo{person}{Tobias
  Hey}, \bibinfo{person}{Martin Armbruster}, \bibinfo{person}{Sophie Corallo},
  \bibinfo{person}{Dominik Fuchß}, \bibinfo{person}{Jan Keim},
  \bibinfo{person}{Haoyu Liu}, {and} \bibinfo{person}{Anne Koziolek}.}
  \bibinfo{year}{2025}\natexlab{}.
\newblock \bibinfo{title}{Software Architecture Meets LLMs: A Systematic
  Literature Review}.
\newblock
\showeprint[arxiv]{2505.16697}~[cs.SE]
\urldef\tempurl%
\url{https://arxiv.org/abs/2505.16697}
\showURL{%
\tempurl}


\bibitem[Schoenfeld(2016)]%
        {Schoenfeld:2016}
\bibfield{author}{\bibinfo{person}{Alan~H. Schoenfeld}.}
  \bibinfo{year}{2016}\natexlab{}.
\newblock \showarticletitle{Learning to Think Mathematically: Problem Solving,
  Metacognition, and Sense Making in Mathematics (Reprint)}.
\newblock \bibinfo{journal}{\emph{Journal of Education}}  \bibinfo{volume}{196}
  (\bibinfo{date}{4} \bibinfo{year}{2016}), \bibinfo{pages}{1--38}.
\newblock
Issue 2.
\showISSN{0022-0574}
\href{https://doi.org/10.1177/002205741619600202}{doi:\nolinkurl{10.1177/002205741619600202}}


\bibitem[Sengul et~al\mbox{.}(2024)]%
        {Sengul:2024}
\bibfield{author}{\bibinfo{person}{Cigdem Sengul}, \bibinfo{person}{Rumyana
  Neykova}, {and} \bibinfo{person}{Giuseppe Destefanis}.}
  \bibinfo{year}{2024}\natexlab{}.
\newblock \showarticletitle{Software engineering education in the era of
  conversational AI: current trends and future directions}.
\newblock \bibinfo{journal}{\emph{Frontiers in Artificial Intelligence}}
  \bibinfo{volume}{7} (\bibinfo{date}{8} \bibinfo{year}{2024}).
\newblock
\showISSN{2624-8212}
\href{https://doi.org/10.3389/frai.2024.1436350}{doi:\nolinkurl{10.3389/frai.2024.1436350}}


\bibitem[Sharma and Wallace(2025)]%
        {Sharma:2025}
\bibfield{author}{\bibinfo{person}{Ansh Sharma} {and} \bibinfo{person}{James~R
  Wallace}.} \bibinfo{year}{2025}\natexlab{}.
\newblock \showarticletitle{DeTAILS: Deep Thematic Analysis with Iterative LLM
  Support}. In \bibinfo{booktitle}{\emph{Proceedings of the 7th ACM Conference
  on Conversational User Interfaces}} (New York, NY, USA).
  \bibinfo{publisher}{ACM}, \bibinfo{pages}{1--7}.
\newblock
\showISBNx{9798400715273}
\href{https://doi.org/10.1145/3719160.3735657}{doi:\nolinkurl{10.1145/3719160.3735657}}


\bibitem[Stewart and Shamdasani(2015)]%
        {Stewart:2015}
\bibfield{author}{\bibinfo{person}{David~W. Stewart} {and}
  \bibinfo{person}{Prem~N. Shamdasani}.} \bibinfo{year}{2015}\natexlab{}.
\newblock \bibinfo{booktitle}{\emph{Focus Groups: Theory and Practice}
  (\bibinfo{edition}{3} ed.)}.
\newblock \bibinfo{publisher}{Sage}, \bibinfo{address}{London}.
\newblock


\bibitem[Storey(2026)]%
        {Storey:2026}
\bibfield{author}{\bibinfo{person}{Margaret-Anne Storey}.}
  \bibinfo{year}{2026}\natexlab{}.
\newblock \showarticletitle{From technical debt to cognitive and intent debt}.
\newblock \bibinfo{journal}{\emph{ACM Queue}} \bibinfo{volume}{24},
  \bibinfo{number}{2} (\bibinfo{year}{2026}).
\newblock


\bibitem[Sweller(1994)]%
        {Sweller:1994}
\bibfield{author}{\bibinfo{person}{John Sweller}.}
  \bibinfo{year}{1994}\natexlab{}.
\newblock \showarticletitle{Cognitive load theory, learning difficulty, and
  instructional design}.
\newblock \bibinfo{journal}{\emph{Learning and instruction}}
  \bibinfo{volume}{4}, \bibinfo{number}{4} (\bibinfo{year}{1994}),
  \bibinfo{pages}{295--312}.
\newblock


\bibitem[Ullrich et~al\mbox{.}(2025)]%
        {Ullrich:2025}
\bibfield{author}{\bibinfo{person}{Jonathan Ullrich}, \bibinfo{person}{Matthias
  Koch}, {and} \bibinfo{person}{Andreas Vogelsang}.}
  \bibinfo{year}{2025}\natexlab{}.
\newblock \bibinfo{title}{From Requirements to Code: Understanding Developer
  Practices in LLM-Assisted Software Engineering}.
\newblock
\showeprint[arxiv]{2507.07548}~[cs.SE]
\urldef\tempurl%
\url{https://arxiv.org/abs/2507.07548}
\showURL{%
\tempurl}


\bibitem[Vella and Blincoe(2026)]%
        {Vella:2026:Longitudinal}
\bibfield{author}{\bibinfo{person}{Annie Vella} {and} \bibinfo{person}{Kelly
  Blincoe}.} \bibinfo{year}{2026}\natexlab{}.
\newblock \bibinfo{title}{The Impact of AI Coding Assistants on Software
  Engineering: A Longitudinal Study}.
\newblock
\showeprint[arxiv]{2605.23135}~[cs.SE]
\urldef\tempurl%
\url{https://arxiv.org/abs/2605.23135}
\showURL{%
\tempurl}


\bibitem[Wang et~al\mbox{.}(2026)]%
        {Wang:2026}
\bibfield{author}{\bibinfo{person}{Yingying Wang}, \bibinfo{person}{Masih~Beigi
  Rizi}, \bibinfo{person}{Fatemeh Khashei}, {and} \bibinfo{person}{Julia
  Rubin}.} \bibinfo{year}{2026}\natexlab{}.
\newblock \showarticletitle{From Specifications to Implementation in the Gen-AI
  Era: Lessons from a Project-Based Software Engineering Course}.
\newblock \bibinfo{journal}{\emph{Proceedings of the ACM on Software
  Engineering}}  \bibinfo{volume}{3} (\bibinfo{date}{6} \bibinfo{year}{2026}),
  \bibinfo{pages}{1424--1447}.
\newblock
Issue FSE.
\showISSN{2994-970X}
\href{https://doi.org/10.1145/3797092}{doi:\nolinkurl{10.1145/3797092}}


\bibitem[Wang and Chuang(2024)]%
        {Wang:2024}
\bibfield{author}{\bibinfo{person}{Yu-Yin Wang} {and} \bibinfo{person}{Yu-Wei
  Chuang}.} \bibinfo{year}{2024}\natexlab{}.
\newblock \showarticletitle{Artificial intelligence self-efficacy: Scale
  development and validation}.
\newblock \bibinfo{journal}{\emph{Education and Information Technologies}}
  \bibinfo{volume}{29} (\bibinfo{date}{3} \bibinfo{year}{2024}),
  \bibinfo{pages}{4785--4808}.
\newblock
Issue 4.
\showISSN{1360-2357}
\href{https://doi.org/10.1007/s10639-023-12015-w}{doi:\nolinkurl{10.1007/s10639-023-12015-w}}


\bibitem[Wohlin and Rainer(2022)]%
        {Wohlin:2022}
\bibfield{author}{\bibinfo{person}{Claes Wohlin} {and} \bibinfo{person}{Austen
  Rainer}.} \bibinfo{year}{2022}\natexlab{}.
\newblock \showarticletitle{Is it a case study?—A critical analysis and
  guidance}.
\newblock \bibinfo{journal}{\emph{Journal of Systems and Software}}
  \bibinfo{volume}{192} (\bibinfo{date}{10} \bibinfo{year}{2022}),
  \bibinfo{pages}{111395}.
\newblock
\showISSN{01641212}
\href{https://doi.org/10.1016/j.jss.2022.111395}{doi:\nolinkurl{10.1016/j.jss.2022.111395}}


\bibitem[Zhao and Tsuboi(2024)]%
        {Xin:2024}
\bibfield{author}{\bibinfo{person}{Xin Zhao} {and} \bibinfo{person}{Narissa
  Tsuboi}.} \bibinfo{year}{2024}\natexlab{}.
\newblock \showarticletitle{Early Career Software Developers - Are You Sinking
  or Swimming?}. In \bibinfo{booktitle}{\emph{Proceedings of the 46th
  International Conference on Software Engineering: Software Engineering in
  Society}} (New York, NY, USA). \bibinfo{publisher}{ACM},
  \bibinfo{pages}{166--176}.
\newblock
\showISBNx{9798400704994}
\href{https://doi.org/10.1145/3639475.3640106}{doi:\nolinkurl{10.1145/3639475.3640106}}


\bibitem[Zhou et~al\mbox{.}(2024)]%
        {Zhou:2024}
\bibfield{author}{\bibinfo{person}{Kaitlyn Zhou}, \bibinfo{person}{Jena Hwang},
  \bibinfo{person}{Xiang Ren}, {and} \bibinfo{person}{Maarten Sap}.}
  \bibinfo{year}{2024}\natexlab{}.
\newblock \showarticletitle{Relying on the Unreliable: The Impact of Language
  Models’ Reluctance to Express Uncertainty}. In
  \bibinfo{booktitle}{\emph{Proceedings of the 62nd Annual Meeting of the
  Association for Computational Linguistics (Volume 1: Long Papers)}}
  (Stroudsburg, PA, USA). \bibinfo{publisher}{Association for Computational
  Linguistics}, \bibinfo{pages}{3623--3643}.
\newblock
\href{https://doi.org/10.18653/v1/2024.acl-long.198}{doi:\nolinkurl{10.18653/v1/2024.acl-long.198}}


\end{thebibliography}
